\documentclass[twocolumn,tighten,twocolappendix]{aastex63}
\shorttitle{Milky Way's Vestigial Nuclear Jet}
\shortauthors{Cecil~et.~al.}

\newcommand\sgA{Sgr~A$^*$}
\newcommand\kms{~km~s$^{-1}$}
\newcommand\ergs{erg~s$^{-1}$}
\newcommand\pA{Paschen~$\alpha$}
\newcommand\ranA{73\arcdeg--126\arcdeg}
\newcommand\Msun{\nom{M}\ }
\newcommand{\Pjet}{P_\mathrm{jet}}
\newcommand{\rjet}{r_\mathrm{jet}}

\newcommand{\nav}{\bar{n}}
\newcommand{\Nav}{\bar{N}}
\newcommand{\tconf}{t_\mathrm{conf}}
\newcommand{\rmmm}{R_\mathrm{m}}
\newcommand{\ppcc}{\,\mathrm{particles~cm^{-3}}}
\newcommand{\ppcs}{\,\mathrm{cm}^{-2}}

\received{Jun 21, 2021}
\accepted{Aug 26, 2021}
\published{ApJ Dec 6, 2021}
\reportnum{doi.org/10.3847/1538-4357/ac224f}
\begin{document}

\title{Tracing the Milky Way's Vestigial Nuclear Jet}

\correspondingauthor{Gerald Cecil}
\email{cecil@unc.edu}

\author[0000-0003-1346-8481]{Gerald Cecil}
\affiliation{Dept. Physics and Astronomy, University of North Carolina, Chapel Hill, NC 27599, USA}
    
\author[0000-0002-5104-6434]{Alexander Y. Wagner}
\affiliation{Center for Computational Sciences, University of Tsukuba, 1-1-1 Tsukuba, Ibaraki 305-8577, Japan}

\author[0000-0001-7516-4016]{Joss Bland-Hawthorn}{}
\affiliation{Sydney Institute for Astronomy, School of Physics A28,
  University of Sydney, NSW 2006, Australia}
\affiliation{ARC Centre of Excellence for All Sky Astrophysics in 3 Dimensions (ASTRO-3D), Australia} 
  
\author[0000-0003-0234-7940]{Geoffrey V. Bicknell}
\affiliation{Research School of Astronomy \& Astrophysics, Australian National University, Canberra 2611, Australia}

\author[0000-0003-0632-1000]{Dipanjan Mukherjee}
\affiliation{Inter-University Centre for Astronomy and Astrophysics (IUCAA), Post Bag 4, Pune-41007, India}

\begin{abstract}
MeerKAT radio continuum and XMM-Newton X-ray images recently revealed a spectacular bipolar channel at the Galactic Center that spans several degrees ($\sim 0.5$ kpc).
An intermittent jet likely formed this channel and is consistent with earlier evidence of a sustained, Seyfert-level outburst fueled by black hole accretion onto \sgA\ several Myr ago.
Therefore, to trace a now weak jet that perhaps penetrated, deflected, and percolated along multiple paths through the interstellar medium, relevant interactions are identified and quantified in archival X-ray images, Hubble Space Telescope \pA\ images and Atacama Large Millimeter/submillimeter Array millimeter-wave spectra, and new SOAR telescope IR spectra.
Hydrodynamical simulations are used to show how a nuclear jet can explain these structures and inflate the ROSAT/eROSITA X-ray and Fermi $\gamma$-ray bubbles that extend $\pm$75$^\circ$ from the Galactic plane.
Thus, our Galactic outflow has features in common with energetic, jet-driven structures in the prototypical Seyfert galaxy NGC 1068.
\end{abstract}


\section{Introduction}\label{sec:intro}

In recent years we have come to recognize that the \object{Galactic Center} (GC) is recovering from a major power surge.
Due to rapid improvements in multiwavelength spectral imaging, we can now associate the lingering record of this ferocity with structures in classical active galactic nuclei (AGNs).
In particular, radio continuum \citep[MeerKAT,][]{2019Natur.573..235H} and X-ray \citep{2019Natur.567..347P} ridges and shells ever more distant from the Galactic plane resemble energetic outflows from some Seyfert galaxies.
While some of the GC's low-latitude loops may be nearby \citep[e.g.,][]{2020PASJ...72L..10T}, X-rays arch far out of the plane (ROSAT, \citealt{2003ApJ...582..246B}; eROSITA, \citealt{2020Nat.588.227}) to envelop both $\gamma$-ray \citep[``Fermi",][]{2010ApJ...724.1044S} bubbles across $\sim$150\arcdeg\ of sky.

To assess the impacts of such outbursts on the assembly and evolution of an $L^\star$ galaxy like the Milky Way (MW), we must understand how often they occur and which processes power them.
The date and duration of the most recently ended episode of significant power are key.
Quantifying vestigial outcomes such as a fading jet today could probe the pressure gradient in hot gas in the GC, in turn uniquely constraining the duty cycle, accretion flow, and duration of recent AGN episodes in a substantial spiral galaxy.

Set against an ancient stellar population at the GC is a distinct population of young, intermediate mass stars on elliptic orbits. All of these are 3--8 Myr old \citep{2010ApJ...725..188M,2006ApJ...643.1011P,2014ApJ...783..131Y} with several ejected to high velocities within the last 5 Myr \citep{2020MNRAS.491.2465K}.
As such, they likely formed from the same fuel that rejuvenated the nuclear activity \citep{2013MNRAS.433..353L}.

However, the implied rate to form stars of $<0.1$ \Msun\ yr$^{-1}$ is orders of magnitude too slow to inflate structures seen on 10 pc scale let alone a thousand times larger.
Nor can starbursts explain coherent, large-scale patterns in UV absorption lines \citep{2014ApJ...787..147F}, H$\alpha$ \citep{2020ApJ...899L..11K} and X-ray spectral \citep{2016ApJ...829....9M}, H~I \citep{2016ApJ...826..215L} velocities, and the lingering non-stellar ionization of transceivers in the halo \citep{2019ApJ...886...45B}.

These ``fading echoes'' tell us that \sgA at the very center was a thousandfold brighter than its current mean up to a century ago \citep{2017MNRAS.465...45C} and for at least several centuries prior, $10^5$ times brighter several thousand years ago, to a million times brighter within the last Myr \citep[e.g.,][]{2013ASSP...34..331P,2013ApJ...773...20N}, timescales far too brief for stellar evolutionary processes. 
Thus, the consensus that $\sim10^{56-57}$ erg were injected into the interstellar medium (ISM) 2--8 Myr ago \citep[e.g.,][]{2016ApJ...829....9M}, possibly by a powerful jet sustained for up to 1 Myr \citep{2012ApJ...756..181G,2012ApJ...761..185Y,2020ApJ...894..117Z,2021arXiv210903834M}.

Are residual effects of an episodic MW jet evident?
No characteristic, linear radio jet has been imaged at small radii in any dataset or, definitely, on any scale.
But kiloparsec-scale jets persist in otherwise normal barred spiral galaxies \citep[e.g.,][]{2002ApJ...568..627C,2006AJ....132.2233K}, and some galactic black holes with jets are accreting only 100 times faster than the abstemious \sgA\ \citep[e.g.,][]{2018ApJ...852....4D}.
Moreover, very-long-baseline interferometry closure phases down to millimeter wavelengths \citep[mm-VLBI,][]{2004Sci...304..704B} and the inclusion of co-phased Atacama Large Millimeter/submillimeter Array (ALMA) data to improve imaging \citep{2019ApJ...871...30I} support a consistent interpretation of \sgA\ \citep[e.g.,][]{1993A&A...278L...1F} from radio through IR (and to X-rays during flares), and constrains the angle from our sightline of a plausible magnetized jet \citep{2007MNRAS.379.1519M,2019ApJ...871...30I}.
Infrared VLTI mapping has constrained the orbital inclination of bright flares in the current accretion disk of \sgA\ \citep{2018A&A...618L..10G}, providing us with an axis near which to search on larger scales.
Beyond the 0.5 pc radius dominated by \sgA, perhaps a weak nuclear jet lurks in the complicated GC environment.

Therefore, in Section \ref{sec:bubles} we first summarize relevant structures uncovered or linked together by recent, wide-field, multiwavelength mosaics and by interferometry at very small scales.
Then, in Section 3 we identify aligned X-ray, infrared, and molecular structures that we propose may trace the currently active jet across 10 pc as its effects appear sporadically in various wave bands.
Jet dynamical times at this scale would be several centuries, so in Section \ref{sec:sims} our 3D hydrodynamical simulations explore how the cocoon of a low-power, inclined jet might alter the kinematics of adjacent cold gas over several millenia.
Because dynamic structures reach hundreds of pc radii, we also simulate gas flows on that scale. 
Section 5 mentions insights from AGN NGC~1068 whose more powerful jets have been redirected by collisions also on that scale.
Position angle (PA) is measured in (R.A., decl.), but otherwise north etc.\ refer to Galactic coordinates; our figures default to north at top, and  ``the jet" is shorthand for ``our proposed jet." 
At the GC 7.9 kpc from us \citep{2020PASJ...72...50V}. $1\arcsec = 0.04$~pc and $1\arcmin = 2.4$~pc.
All velocities are relative to the local standard of rest.

\begin{figure}
    \centering
    \includegraphics[scale=0.26]{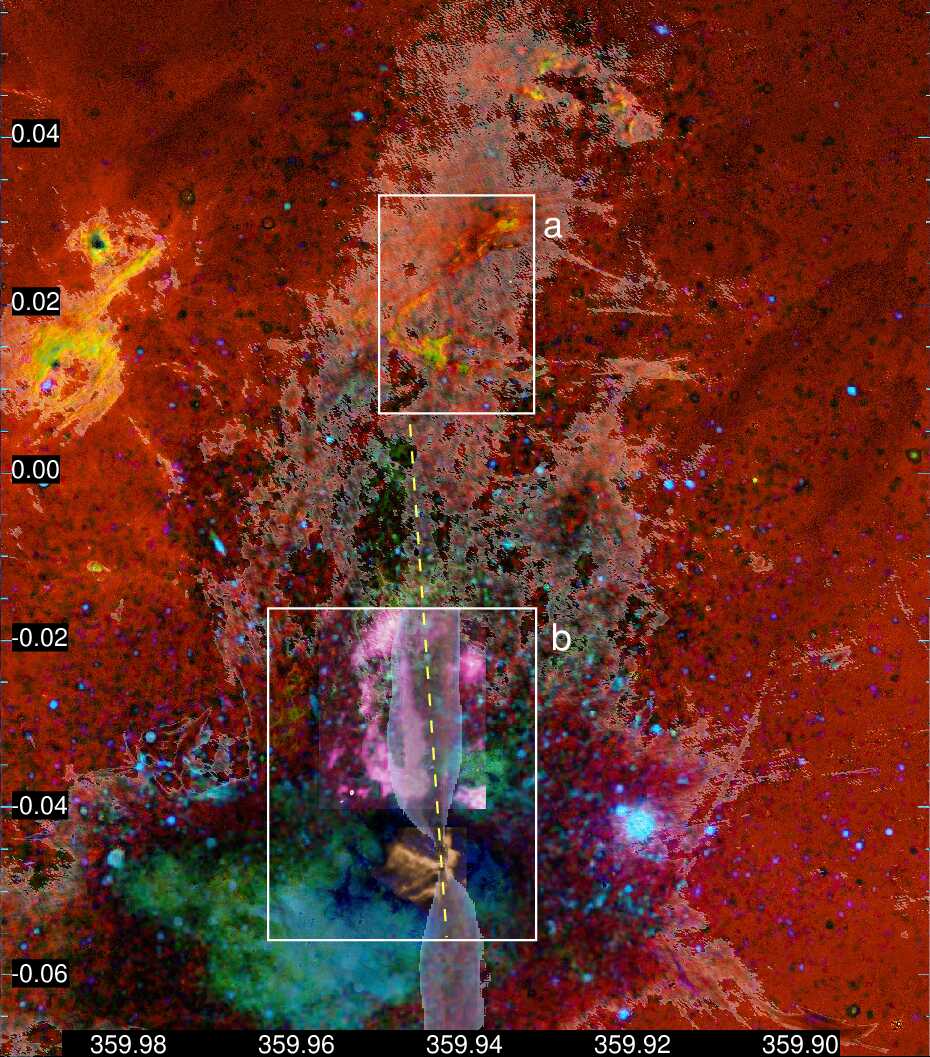}
    \caption{This region at the GC with degree labels shows X-rays, energetic electrons, and molecular gas overlaid on \pA\ from the Hubble Space Telescope (HST) in rust color. The last, shown log-scaled, brightens through orange to yellow especially in the minispiral around \sgA\ in the lower quarter of box b that delineates Figure~\ref{fig:CND} panels (b)+(d). 
The northern X-ray lobe delineated by Chandra in green (2--4.5 keV) and blue (4.5--8 keV, mostly within the CND and at many compact sources) is spanned by radio continuum filaments (ragged gray--white overlay) mapped by the JVLA (shown alone in Figure 2 (e)).
Within box (b) in pink is the redshifted ALMA CS emission discussed in Section 3.3. The translucent overlay plots jet tracer particles from one of our smaller-scale 3D simulations launched along the dashed yellow jet axis. A larger-scale simulation is compared to data in Figures 3 and 31. Box (a) spans Figure 11.}
    \label{fig:ChandraPA}
\end{figure}

\section{Bubbles and Channels}\label{sec:bubles}

Interferometers are imaging the GC in emission lines of various excitation energies and critical densities to isolate kinematical structures now on scales from just outside the innermost stable orbit around the $M_\mathrm{BH}=4.1\times 10^6$ \Msun\ central black hole \citep{2020ApJ...889...61P}, to cooler parts of its accretion disk within 0.01 pc that rotate at $\sim500$\kms\ \citep{2019Natur.570...83M}. Beyond at 0.25~pc lies a possible interaction with one arm of the ionized ``nuclear minispiral" \citep[][LMB13 hereafter]{2013ApJ...779..154L} within the $\sim4.5$ pc diameter molecular Circumnuclear Disk/torus \citep[CND,][T18 hereafter]{2018PASJ...70...85T}.
The CND starts at 1.5~pc radius, tilts $30\pm5$\arcdeg\ from edge-on to us and, critically as we will discuss later, tilts $\sim20$\arcdeg\ to the Galactic plane \citep[Z09 hereafter; T18]{2017ApJ...847....3H,2009ApJ...699..186Z}.
Simulations \citep[e.g.,][]{2020MNRAS.499.4455T} suggest that it is replenished by stellar feedback at rate $\sim0.03$ \Msun yr$^{-1}$. 

Beyond the CND, X-ray emission concentrates into a pair of $5\times10$ pc ellipsoidal lobes elongated from the Galactic plane.
The northern one (Figure~\ref{fig:ChandraPA}) is transected by radial streams of radio continuum emission \citep[Z16 hereafter]{2016ApJ...817..171Z} and dark, radial dust streaks.
Highly ionized edges of many molecular clouds within $5\arcmin$ radius show the fluorescent 6.4 keV iron K$\alpha$ line and reflected continua plus the low-energy end of a plausible Compton bump \citep{2018PASJ...70R...1K} that together signify nonequilibrium photoionization.
Time-correlated variations of their fluxes have been sequenced in space \citep[e.g.,][]{2017MNRAS.465...45C} to trace past eruptions associated with enhanced accretion to infer that \sgA\ powered down within the last century from a state $\sim10^4\times$ brighter than today that had persisted for centuries.

\begin{figure*}
\centering
\includegraphics[scale=0.6]{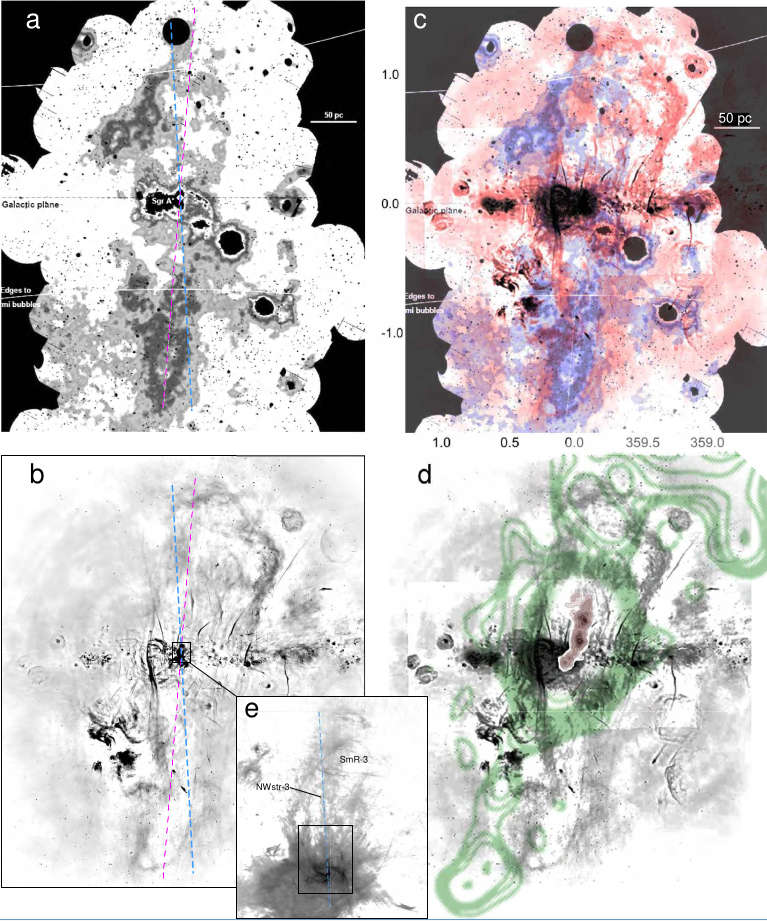}
\caption{(a) \citet{2019Natur.567..347P} Figure~4 in inverse grayscale (1.5--2.6~keV X-rays from XMM-Newton) is blue in (c) where it is overlaid with the MeerKAT 1.28~GHz image (b) from \citet{2019Natur.573..235H} Figure 1 in pink.
Galactic coordinate degrees are marked in (c).
Panel (d) adds to GC structures in (b): green contours show 15 GHz emission in the extended narrow-line region of the nearby AGN NGC~1068 \citep[Very Large Array (VLA),][Figure 1, used with permission of the AAS]{1987ApJ...319..105W} and 5 GHz \citep[MERLIN,][Figure 1, used with permission of the AAS]{2004ApJ...613..794G} beige contours near the nucleus.
The northern radio lobe of NGC 1068 and the northern Fermi Bubble of the MW (arc in (a)) flare out at an identical scale.
The central region at 5.5~Ghz from Z16 is shown with log-scaled intensities in (e), which covers the region plotted in Figure~1; the jet axis in blue is extrapolated on the other panels. The box in (e) delineates Figure~\ref{fig:CND} panels (b)+(d).}
\label{fig:radioXray}
\end{figure*}

The northern X-ray lobe ends in projection on a ``molecular loop" (ML) comprised of two distinct kinematical systems over 100--180 pc in longitude between latitudes $b=0\fdg0-0.\fdg35$ whose profiles in various molecular lines split at -35 and +70 \kms\ centers \citep[Figure~9 of][]{2016MNRAS.457.2675H} near longitude 0\arcdeg.
Also rooted here is one end of the radio continuum ``$\Omega$-lobe" \citep{1984Natur.310..568S} that rises to 15 pc and entrains dust \citep{2003ApJ...582..246B} and ions that (Fig.~\ref{fig:radioXray}) are enveloped by ever larger radio and X-ray arcs inclined by $\sim15^\circ$ toward the northwest to end in similar prows at similar radius; this tilt persists to the largest scale of radio emission (purple dashed line in Figure~\ref{fig:radioXray}).
To the north, the X-ray pattern offsets east of similar radio continuum emission \citep{2019Natur.573..235H,2021A&A...646A..66P}.
The enigmatic, vertical bundle of flux tubes \citep{2021ApJ...920....6G} at a 15\arcmin\ radius east may confine the X-ray emission, which at least in the north protrudes farther eastward once the filaments end. 
To the south, radio synchrotron 1--2~Myr old appears to envelop the X-rays \citep{2013ApJ...773...20N,2013ASSP...34..331P}.
Just those electrons attest to an eruption of $\ga7\times10^{52}$ erg \citep{2021A&A...646A..66P,2019Natur.573..235H}, but certainly orders of magnitude more energetic than that because the entrained ions and dust, being inefficient radiators, retain much more energy.

Outflow up to 14 kpc revealed by eROSITA \citep{2020Nat.588.227} envelops the 50\arcdeg\ (6 kpc) high Fermi $\gamma$-ray bubbles \citep[FB hereafter,][]{2014ApJ...793...64A}
that jet models \citep{2020ApJ...894..117Z}, among other processes, can explain.
The FB have an otherwise constant surface brightness but sport a bright $\gamma$-ray ``cap" at its north boundary and an X-ray bright ``claw" at its southwestern boundary.
Connecting these, an apparent linear ``channel" inclined $15^\circ$ \citep{2012ApJ...753...61S} from the Galactic polar axis that might trace a jet did not survive more Fermi data \citep{2014ApJ...793...64A}, although subtracting fore/background $\gamma$-ray emission is model specific.

Ionized segments of the Magellanic Stream near the southern Galactic pole are our most distant transceivers of past activity. Their unusually high excitation and kinematics \citep{2020ApJ...897...23F} may arise from exposure to a cone of hard-spectrum photons during a prolonged Seyfert phase of the MW 2.5--4.5 Myr ago
\citep{2003ApJ...582..246B,2019ApJ...886...45B}.

\section{Uncovering the Milky Way's Jet}\label{sec:back}
As LMB13 noted, within the GC all candidate jets in various wave bands have quite different position angles (PA hereafter) on sky.
But in Section \ref{sec:Sjet} and in interactive Figure 3 we describe unique features that straddle \sgA\ and may trace the effects of a plausible jet cocoon for 4\farcm2.
LMB13 proposed the one currently active south of \sgA.
Here, we assess three more diametrically across \sgA\ to the north.
Gas moving radially near the GC can be on either side of the CND depending on preference for in- or outflow.
T18 interpreted as inflowing some of the structures that in the following discussion we interpret as outflowing.

\subsection{Jet Near Our Sightline At \sgA?}\label{sec:VLBI}
Especially parallel to the Galactic plane, VLBI views of activity very nearby \sgA\ are blurred by electron scattering that increases as $\lambda^2$.
Millimeter-VLBI arrays blur less to probe closer to \sgA\ where relativistic asymmetries ambiguate separation of accretion flows from jet flows, especially if the jet axis is not near the sky plane or if it starts beyond the scale accessed by the shorter baselines of a sparse antenna array.

Closure phases at $\lambda$7 mm fitted by \citet{2007MNRAS.379.1519M} to their model of a bipolar, freely expanding jet radiating synchrotron yielded a jet axis along PA~90\arcdeg--120\arcdeg, $<13$\arcdeg\ on sky from the Galactic polar axis, and inclined $\ga75$\arcdeg\ to our sightline.
Symmetric closure phases at $\lambda$3.5 mm taken with sparse north-south baselines \citep{2011ApJ...735..110B} allowed a 48\arcdeg--73\arcdeg\ inclination to avoid Doppler boosting one side of a bipolar jet.

Subsequently, with co-phased ALMA, LMT, and the GBT linked in too, \citet{2019ApJ...871...30I} found negligible asymmetry of the intrinsic (i.e., unscattered) image of \sgA, hence that any jet must launch $\la20$\arcdeg\ from our sightline (Fig.~\ref{fig:orbitErrors}(b) circled insert).
This is consistent with the 15\arcdeg$\pm$10\arcdeg\ inclination of \citet{2019Natur.570...83M} to the broad, double-peaked H30$\alpha$ line profiles of the cooler part of the accretion disk. That transition must be maser amplified to reconcile with the narrow Br~$\gamma$ line profile here \citep{2021ApJ...910..143C}.
Likewise, VLTI polarimetric astrometry of hot spots by the \citet{2018A&A...618L..10G} has oriented the inner edge of the accretion disk to $<27$\arcdeg\ from face-on and the spin axis along PA~25\arcdeg--70\arcdeg\ so that a jet would incline 10\arcdeg--43\arcdeg\ from the Galactic plane.
Improved polarimetric mm-VLBI and VLTI imaging of flares will better constrain the disk's aspect, the launch radius of any jet, and perhaps any structure due to jet deflection.

In the rest of this section we delineate relevant structures from prior work then present our new insights on increasing spatial scales.

\begin{figure}
%\begin{interactive}{js}{FigSet3.zip}
\centering
\includegraphics[scale=0.13]{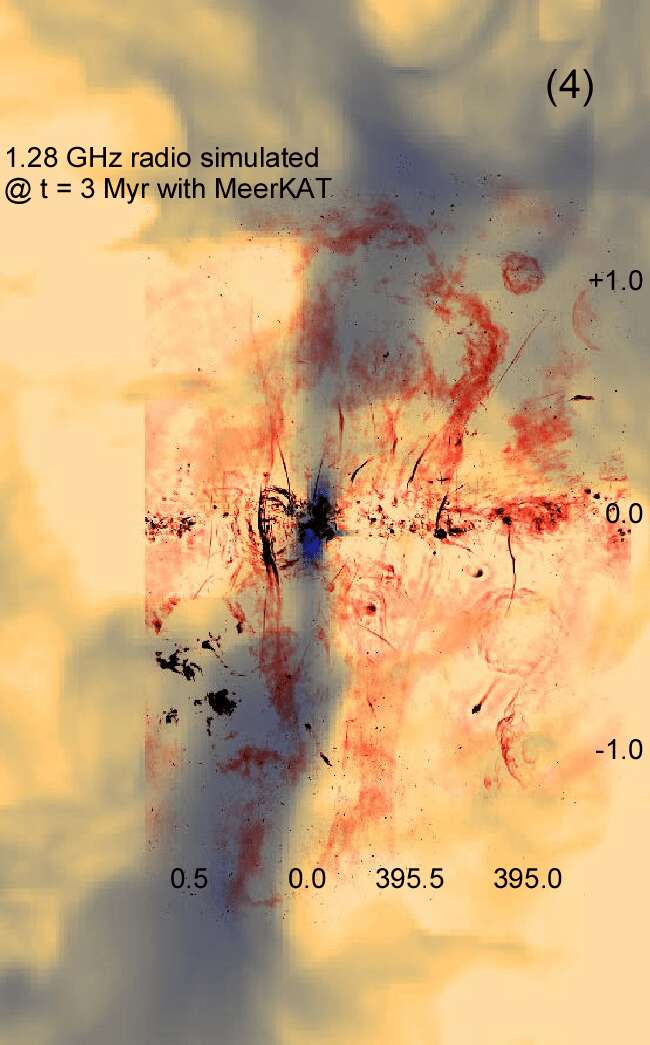}
%\end{interactive}
\caption{This interactive figure online shows how the multi-frequency GC datasets and the resulting structures discussed relate spatially. This example compares the 1.28 GHz MeerKAT (Figure~\ref{fig:radioXray}(b) image (red) from \citet{2019Natur.573..235H} to the derived radio continuum from a kiloparsec-scale simulation discussed in Section 4.1. Right/left arrow keys navigate through the following 17 registered images.
(1) 1.28 GHz MeerKAT; (2) 1.5--2.6 keV XMM-Newton in inverse grayscale (P19); (3) 1.28 GHz in red, 1.5--2.6 keV in blue (P19); (4) 1.28 GHz MeerKAT and radio simulated at \textit{t} = 3 Myr; (5) 1.5--2.6 keV XMM-Newton (P19), with closeup \citet{2015MNRAS.453..172P} Figure~11 whose caption explains the colors; (6) 1.5--2.6 keV XMM-Newton closeup with Chandra 2.6--4.5 keV superimposed in inverse grayscale and the CND delineated by the ellipse; (7) Chandra 2.6--4.5 keV full resolution atop 4.5--8 keV XMM-Newton (P19); (8) 5.5 GHz JVLA from \citet{2016ApJ...817..171Z} with the CND delineated; (9) Closeup of 5.5 GHz JVLA log intensity-scaled with CND delineated; (10) Paschen-$\alpha$ HST/NICMOS from \citet{2010ApJ...709...27W}; (11) Chandra 2.6--4.5 keV red, 4.5--8 keV blue (P19); (12) 5.5 GHz JVLA with Chandra 2.6--8 keV (P19) showing the southern X-ray jet; (13) ALMA CS +77.5 km/s channel \citep{2018PASJ...70...85T}, H42$\alpha$ in orange \citep{2017ApJ...842...94T}. The counter-jet is projected northward from the southern X-ray jet. (14) Simulated jet particle density at $t$ = 3.6 kyr superimposed on the T18 ALMA CS +77.5 \kms\ datacube slice; (15) T18 datacube slice and our inferred geometry; (16) Chandra 2.4--8 keV magenta-green-blue from NASA/CXC/STScI Wang (press release STScI-2009-28); (17) Paschen-$\alpha$ HST/NICMOS atop the T18 ALMA datacube slice.}
\end{figure}

\begin{figure}[b]
    \centering
    \includegraphics[scale=0.37]{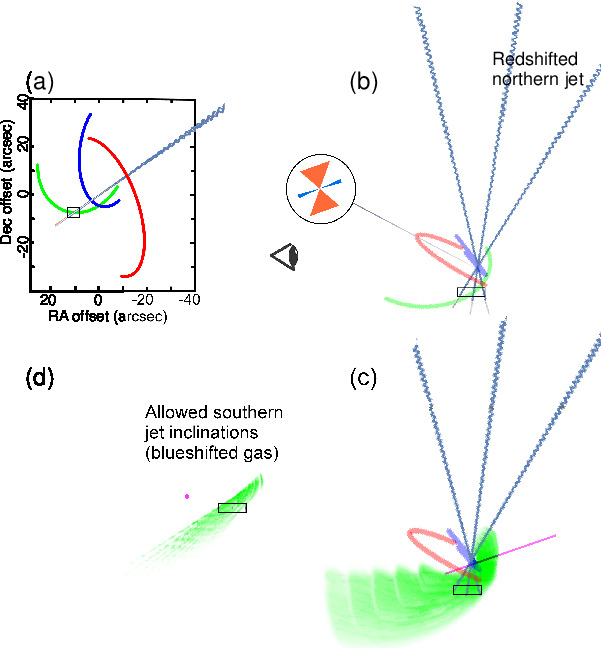}
    \caption{Keplerian orbits of the ionized minispiral and orientations of a bipolar jet anchored at their collision south of \sgA\ as delineated by boxes. The Western Arc (red) sits in the midplane of the CND. (a) Our view of the mean orbits oriented in (R.A., decl.). (b-d) Views are rotated to Galactic coordinates on the same scale as in (a) then revolved 90\arcdeg\ to our vantage at left as shown in (b).
    (c) The $\pm1\sigma$ spread of the allowed Eastern Arm (green) orbits along our sightline. Panel (d) restricts orbits in (c) to those that intersect the southern interaction, repeating this box in (c) and (b); hence, we orient the jet along our sightline. Note that two-thirds of the allowed range redshifts a northern jet away from us.
    In panel (b) the thick accretion disk (red, and perpendicular jet blue) is magnified in cross-section to orient it along the \citet{2018A&A...618L..10G} and \citet{2019ApJ...871...30I} angle; magenta line and dot in (c) and (d) project this jet axis to miss the collision site.}
    \label{fig:orbitErrors}
\end{figure}

\subsection{Jet Orientations Within 1 pc South of \sgA}\label{sec:Sjet}

\subsubsection{Prior work}

\begin{figure*}
    \centering
    \includegraphics[scale=0.61]{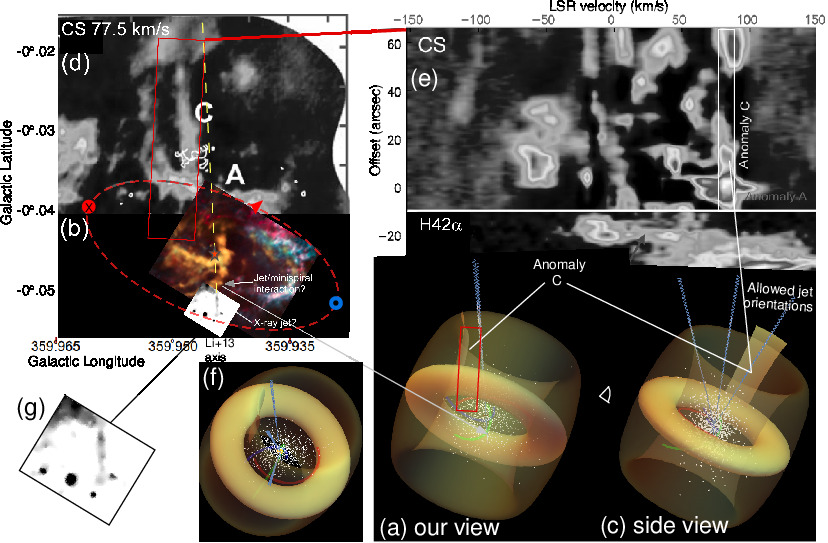}
\caption{Jet orientations that we infer within 3.5 pc of \sgA.
(a) We view the 1.5--2 pc thick, 2 pc high CND torus tilted relative to the Galactic plane and away from us by $\sim25$\arcdeg.
The red box corresponds to that plotted in panels (d) and (e).
In panels (a-c) and (f) the three-armed ionized ``nuclear minispiral" is plotted using the mean of the Keplerian orbit uncertainty distributions (Z09; \citealt{2017ApJ...842...94T}).
The minispiral and the ionized accretion disk \citep{2019Natur.570...83M} sit within the nuclear star cluster whose hot stars carve out the torus.
(b) Combined radio and mid-infrared images of the minispiral are shown with, at right, the molecular emission colored by its line-of-sight velocity \citep[ALMA Observatory press release image, \url{www.almaobservatory.org/en/audiences/cloudlets-swarm-around-our-local-supermassive-black-hole/} based on][Figure 1]{2018A&A...618A..35G}.
The red ellipse delineates the rotating CND; Z09 show that its top half is the near side of the cavity.
South of \sgA\ LMB13 proposed a jet interaction with the Horizontal/Eastern Arm of the minispiral and its putative linear X-ray extension shown here in inverse grayscale in the white box (which is magnified in panel (g)).
The yellow dashed line extends this axis northward.
(c) This view is rotated around the vertical axis of panel (a) by 90\arcdeg\ so we view it from the left-hand edge as drawn.
In our interpretation the jet tilts away from the CND's spin axis for almost all of the range of inclination angles shown whilst intersecting the minispiral Eastern Arm at its uncertain location along our sightline; the jet tilt to us is therefore uncertain by $\sim\pm15$\arcdeg.
Panels (d) and (e) are based on Figure~12 of T18.
(d) ALMA +77.5 \kms\ CS($J=2-1$) channel map  north of \sgA\ with intensities shown in cyclical grayscale (T18).
The yellow dashed line along PA~124\arcdeg\ in panel (b) is extrapolated northward along the southern X-ray jet/\sgA axis.
Note the two parallel linear features offset by 15\arcsec-30\arcsec\ east that run almost north-south for $\sim55\arcsec$. 
We propose that much of this molecular gas is being entrained by the jet.
(e) This position-velocity emission map of CS (above the horizontal line) and recombining H (below) is extracted from the north-south red box in the 77.5\kms\ channel (d); that channel is delineated by the white box.
As discussed in T18, Anomaly C has anomalous velocities relative to the predominantly counter-clockwise rotating CND.
In interpretative panels (a), (c), and (f) redshifted Anomaly C is gas associated with the receding jet on the rear side of the CND and through the periphery of the larger molecular disk within which the inclined CND is embedded.
(f) View down the Galactic north polar axis to show the range of allowed jet trajectories, and its strongest interaction (Anomaly C) drawn shaded near the top left for the range of allowed jet orientations from Figure~\ref{fig:orbitErrors}. We view this panel from its bottom edge.
(g) Magnified X-ray image from panel (b).
\label{fig:CND}}
\end{figure*}

The Chandra X-ray archive provided many calibrated ACIS-I detector event files\footnote{We used sequences 5951, 5950, 4684, 4683, 3549, 3665, 3393, 3392, 3663, 2951, 2943, 2954, 2052, and 242.} of the GC that we merged with CIAO-v4.11 to create an image similar to those of \citet{2003ApJ...591..891B} and \citet{2019ApJ...875...44Z}.
The resulting Figures~\ref{fig:CND}(b) and (g) show source \object{G359.944-0.052}, which is marginally resolved at 0.04 pc across \citep{2019ApJ...875...44Z} and extends 0.3 pc along PA~124\fdg5$\pm$1\fdg5.
LMB13 proposed a synchrotron origin of its featureless spectrum and constrained its properties as a jet. \citet{2019ApJ...875...44Z} found that its initial hard X-ray spectrum softened along its length and hence inferred synchrotron cooling of a few years from a supply of relativistic electrons that has been stable over 20 yr of observation.

\begin{figure}
    \centering
    \includegraphics[scale=0.26]{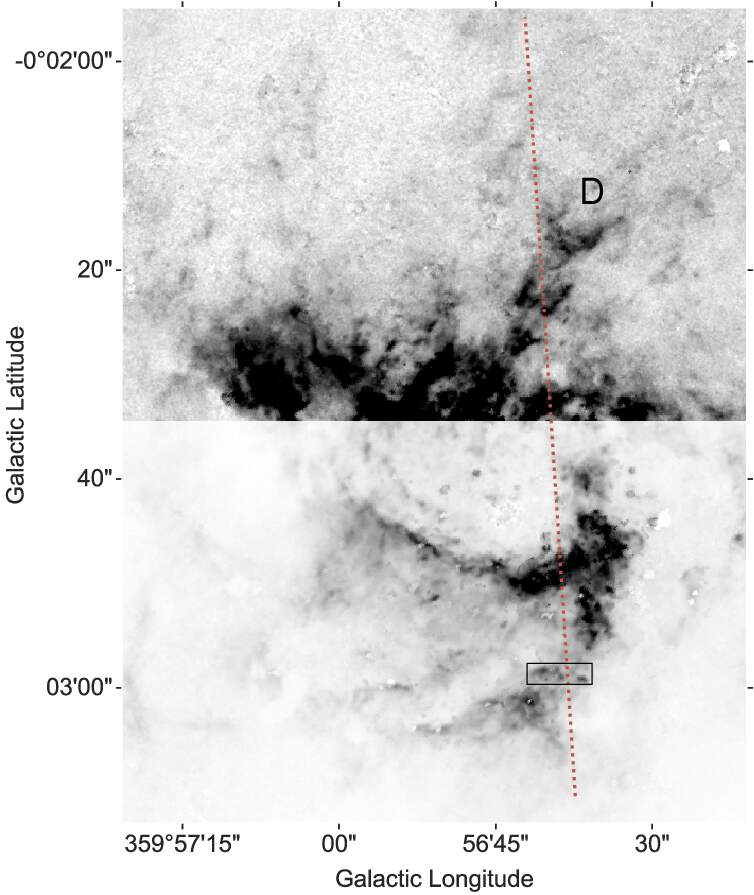}
    \caption{Ionized H near the GC extracted from the NICMOS/HST \pA\ Legacy Survey of the Galactic Centre \citep{2010MNRAS.402..895W}.  The bottom half of the image is rescaled in intensity to 1/8 that of the top to detail the minispiral.
    The bottom of the T18 kinematical Anomaly D is labeled.
    The dotted jet axis is defined by \sgA\ and the interaction with the Eastern (bottom) minispiral Arm within the box magnified in Figure~\ref{fig:nshock} (bottom).}
    \label{fig:PAnearGC}
\end{figure}

Between it and \sgA\ LMB13 also identified the $<$-shaped ``Seagull Nebula" in VLA images, and argued that it is a jet/ISM interaction evident also as a gap in the proximate minispiral arm (see Figure~\ref{fig:PAnearGC} bottom) where adjacent IR spectra show shock-excited emission-line ratios.
Because the nebula is wider than the X-ray feature, they suggested that the nebula spans the lower-momentum jet cocoon.
Summing over the shock front yields a dereddened \pA\ luminosity of $\sim2.3\times10^{30}$ \ergs\ \citep{2010MNRAS.402..895W}.

\begin{figure*}
    \centering
    \includegraphics[scale=0.6]{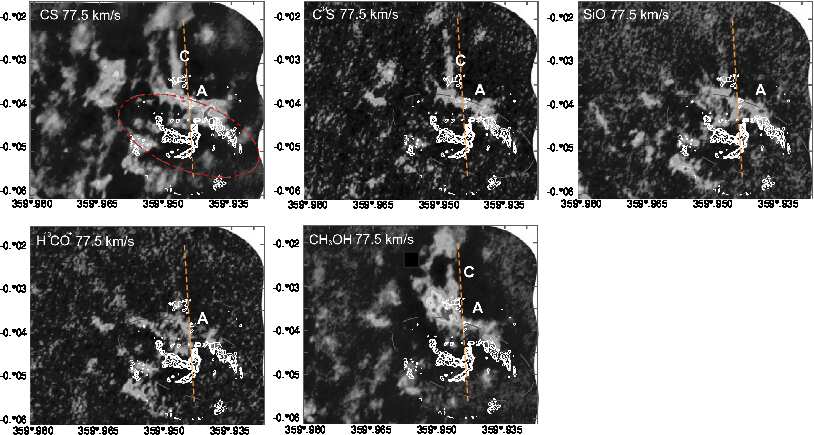}
\caption{ALMA LSR +77.5\kms\ channel in diagnostic lines (based on parts of Figures~3--6 of T18) after continuum subtraction.
The lines show that Anomaly C has $10^3~(T_\mathrm{ex}/200$ K) \Msun\ of weakly shocked molecular gas at $\la10^4$ cm$^{-3}$.
The CND is the dashed ellipse, and the nuclear minispiral within it is shown as white 100 GHz continuum contours.
The dashed line extrapolates the southern jet-\sgA\ axis to the north; it lies slightly west of Anomaly C.
}
\label{fig:anomalyC}
\end{figure*}

\begin{figure*}
    \centering
    \includegraphics[scale=0.46]{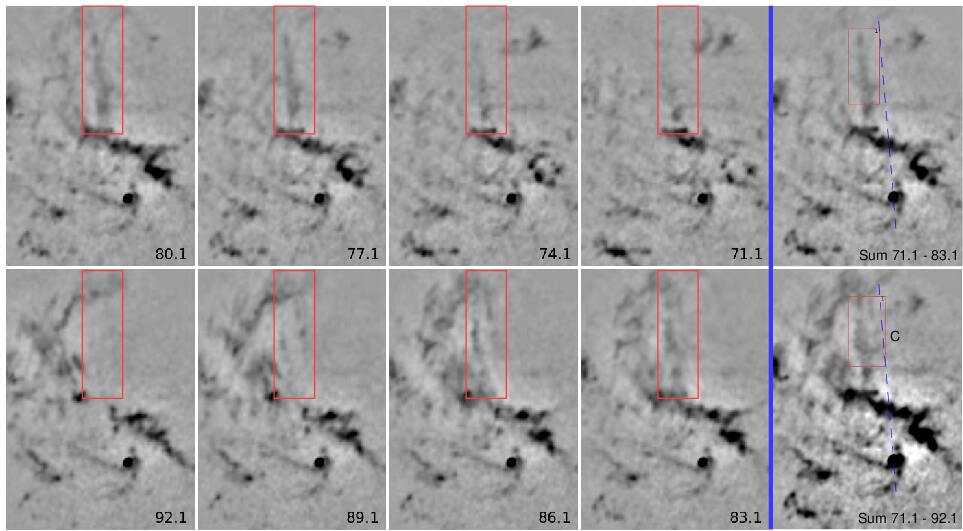}
    \caption{Inverse grayscale image over $1.6\times2.3$ arcmin$^2$ of ALMA CS($J=2-1$) channels, and their sums in the rightmost column; velocity in\kms\ relative to the LSR.
    The region plotted is similar to that in Figure~\ref{fig:CND}(d).
    Continuum emission has not been subtracted but is only noticeable at \sgA\ (black dot) and in the minispiral.
    The proposed jet impact to the north, Anomaly C, is delineated by the red box to show how it shifts from parallel strands starting at bottom left to center filled in the top row.
    The dashed projected jet axis adjacent to the fainter western strand is drawn near the right side within the box. The rightmost column sums emission over the velocities indicated.}
    \label{fig:Ckinematics}
\end{figure*}

Z09 derived 3D orbits from proper+Doppler motions of clumps in the minispiral, found mostly Keplerian motion, and established uncertainties mostly along our sightline (their Table 5); small deviations from Keplerian \citep{2017ApJ...842...94T,2018A&A...618A..35G} imply enough confinement to prevent the thermal dispersal of ionized gas but insufficient to ram decelerate it.
Recent dust polarization maps from SOFIA/HAWC+ indicate some magnetic confinement \citep{2021cosp...43E1251M}, and the twisted morphology of a minispiral arm (Figure~\ref{fig:PAnearGC}) is likewise suggestive.

\subsubsection{New insights on interaction with the CND}

We projected onto our sightline the uncertain Keplerian orbit of the minispiral Eastern Arm (Figure~\ref{fig:orbitErrors}).
If the southern jet is indeed interacting with the Eastern Arm, only the subset of the orbits in panel (c) that remain in panel (d) are allowed by our sightline.
This set orients the jet across the $\sim50$\arcdeg\ range shown in side-view Figure~\ref{fig:CND}(c) and overhead panel (f).
Its axis at this scale would therefore be 1\fdg5--30\arcdeg\ from the Galactic polar axis, and the tilt of the CND torus places the jet axis closer to its eastern end than to its western.

The minispiral interaction therefore inclines the northern jet \ranA\ to us, consistent with jet+scattering models that reproduce the mm-VLBI visibilities of \sgA\ \citep{2007MNRAS.379.1519M} but contrary to recent more face-on determinations (Figure~\ref{fig:orbitErrors}(b) circled insert and magenta line in (c)).

\subsection{Jet Influence Visible North of \sgA\ Within 2 pc}\label{sec:allN}
Here, we examine possible interactions of a counter-jet with the ISM north of \sgA.
None appear in X-rays or radio continua.
However, T18 highlight very suggestive molecular structures.
Their program 2012.1.00080.S archived full data cubes of C$^{34}$S, CH$_3$OH lines, H42$\alpha$, and CS channels LSR 2.5 to +127.5\kms\ (remaining blueshifted channels of CS were flagged for quality control and therefore were unavailable).
\sgA\ lies near the edge of their imaged field using combined 12~m and 7~m dish baselines. Hence, new observations will be needed for features beyond 1\farcm6 north.
Other archived ALMA datasets of the GC are too insensitive or narrow field for us, so in Section \ref{sec:Njet} we extend coverage with a published Nobeyama 45~m dish spectral map.

Recall that northward flow along most of the orientations allowed by a southern jet interaction (Figure~\ref{fig:orbitErrors}) would appear redshifted.
We view the Eastern Arm behind \sgA\ where its uncertain orbit permits interaction with a northern colinear jet for some of these orientations. 

\subsubsection{Anomaly C of T18}

The ALMA data cubes reveal a uniquely linear structure within the otherwise complicated motions of highly excited molecular gas that project on the GC region.
Specifically, T18 identified (their Figure~12) molecular gas having kinematics anomalous to the $\sim115\pm10$\kms\ rotation plus 23$\pm5$\kms\ average contraction of the CND that they derived.
They clarified deviations from coherent ``streamers" that are plausibly associated with fueling \citep[e.g.,][and references therein]{2017ApJ...847....3H}, finding that Anomaly C between 71--92\kms\ has strong CS($J=2-1$), C$^{34}$S, and CH$_3$OH lines but faint/absent in others (Figure~\ref{fig:anomalyC}) including H42$\alpha$.
They used Figure~\ref{fig:CND}(e) to show that this Anomaly is not connected kinematically to the minispiral.

\subsubsection{New insights on Anomaly C} 

The beam taper used by T18 to image the region introduced enough uncertainty in the continuum level (suppressed in the dark areas of Figure~\ref{fig:CND}(e)) to prevent us from extracting reliable emission-line profiles across Anomaly C. Therefore, in Figure~\ref{fig:Ckinematics} we simply sum emission over sequential channel maps.
There, Anomaly C appears as linear features in the 71--92\kms\ channels, so that its velocity dispersion is near the median of values throughout the central molecular zone \citep{2016MNRAS.457.2675H}, which implies Mach numbers $>25$ for this molecular flow.
Northward its redshifted velocities decrease by $<5$\kms\ along its $\sim50$\arcsec\ length.
Its double strands at its highest receding velocities merge together at 83\kms\ for 12\kms, then fade completely by 71\kms.
C$^{34}$S and CH$_3$OH lines trend the same way.

\begin{figure}
    \centering
    \includegraphics[scale=0.25]{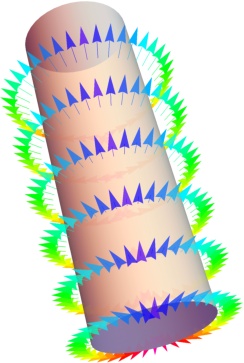}
    \caption{In our interpretation of the molecular line kinematics, the jet cocoon is predominantly expanding radially from the axis of the cylinder, its backside is extinguished, and it is redshifted overall by its tilt away from us. Each vector is colored by its line-of-sight velocity, which goes to 89 \kms\ (green) at the left and right edges of the cylinder.}
    \label{fig:flow}
\end{figure}

This is the coherent pattern of a mostly transverse flow toward us, in an opaque semi-cylinder tilted away from us to provide the overall redshift (Figure~\ref{fig:flow}), and brightened where closest to the inner edge of the CND.
Summing flux over the boxes in the rightmost column of Figure~\ref{fig:anomalyC}, the double strand comprises 27\% of the mass of Anomaly C; the single strand has the rest.

CS and H$^{13}$CO transitions track dense molecular gas at $n$(H$_2) = 10^4$ and 10$^5$ cm$^{-3}$, respectively.
C$^{34}$S/CS tracks an optical depth at $n$(H$_2) = 10^4$ cm$^{-3}$.
CH$_3$OH forms on cold dust whose ice but not grain is evaporated in weak shocks (a jump $\Delta V \la10$\kms);
within the velocities spanned by Anomaly C, it is displaced $\sim3$~\arcsec\ west of the other lines (i.e., toward the projected jet axis).
Thus, Figure~\ref{fig:anomalyC} shows Anomaly C to be weakly (C-)shocked gas at $\la10^4$ cm$^{-3}$.

Assuming LTE, T18 established its total mass as $10^3 (T_\mathrm{ex}/200$ K)\Msun, $\sim3\%$ of the CND's mass.
We used the channel sums shown in the right-hand column of Figure~\ref{fig:Ckinematics} to isolate the western strand closest to the jet in channels 83-92 \kms, which we found comprises 16\% of the mass of Anomaly C hence a kinetic energy $\sim 1.6\times10^{47}(T_\mathrm{ex}/200$ K) erg.
By comparison, the jet seen by \citet{2018Natur.554..334M} to flow from an $\sim15$\Msun\ protostar at $>300$ \kms\ for $\sim$10 pc encompasses $\sim0.1$\Msun\ of ionized gas and could ultimately entrain $<10$\Msun\ of molecular gas to $\sim10$\kms.
While unquestionably the star-forming environment of the GC is unusual, nonetheless to explain Anomaly C without \sgA, more than 100 such intermediate mass protostars would have had to form in the nuclear cluster within the last $\sim40$ kyr and develop a single directional flow without disturbing the rest of the CND.

\subsubsection{Anomaly A of T18}

\begin{figure*}
    \centering
    \includegraphics[scale=0.21]{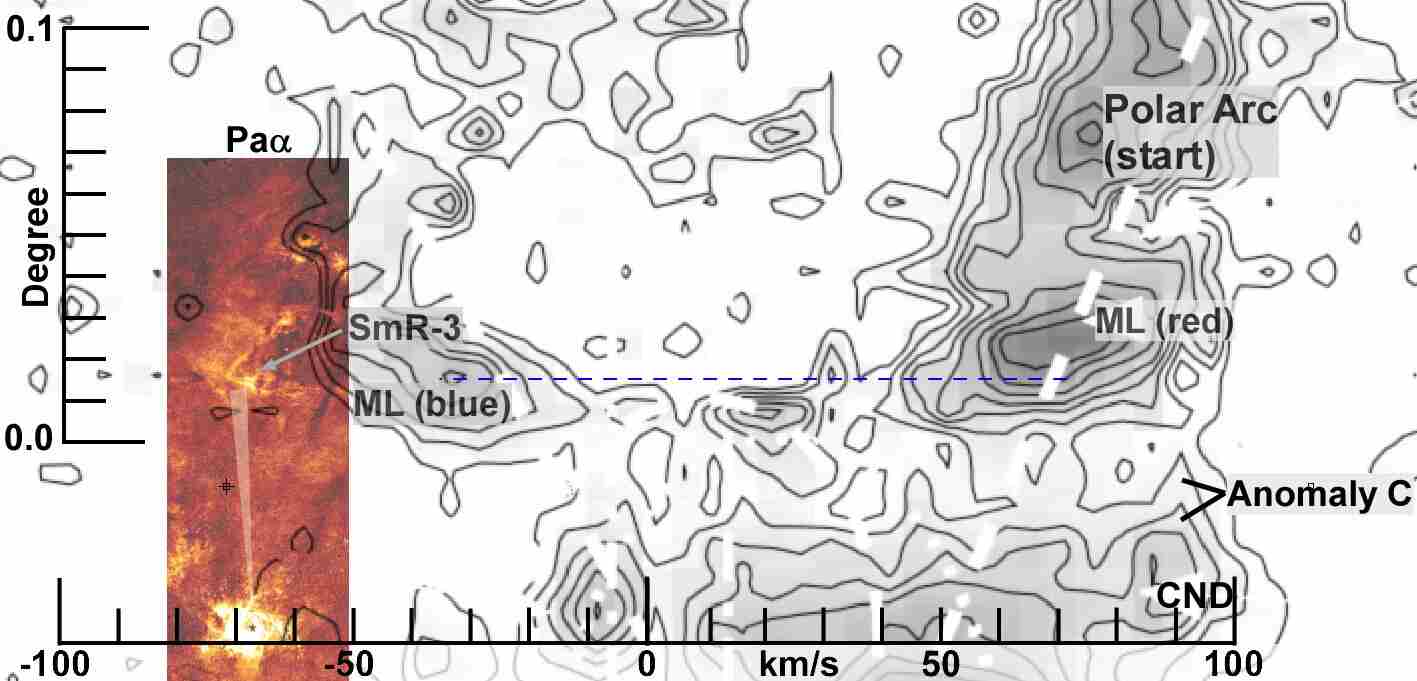}
    \caption{The base image contours latitude-velocity ($b$-$v$) CS($J=1-0$) emission-line profiles (based on Figure~7 of \citealt{2016ApJ...831...72H} region C, reproduced by permission of the AAS) from the Nobeyama 45m dish. The velocity scale is at bottom and latitude is at left.  These velocity profiles are extracted by summing over the Galactic longitudes delineated by the box that superimposes the \pA\ emission in color and shows the interaction in Figure~\ref{fig:Palpha} at the southern edge of the SmR-3 nebula of Z16.
    Note the faint northward extension of the jet/Anomaly C in the red wing of the profiles at 70--90 \kms\ at bottom right; it is not evident in the CS($J=2-1$) PVD of \citet{2016ApJ...831...72H}.  ``ML blue" and ``ML red" are parts of the very extensive molecular features discussed by \citet{2016MNRAS.457.2675H} and references therein.
   }
    \label{fig:Nobfig}
\end{figure*}
\begin{figure*}
    \centering
    \includegraphics[scale=0.32]{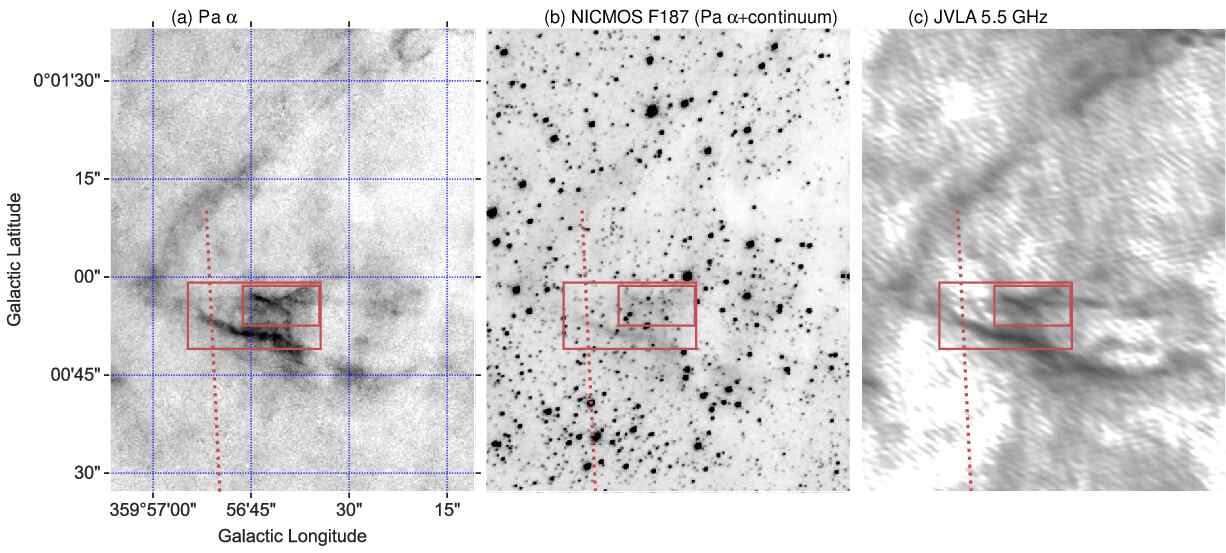}
    \caption{The possible jet/ISM impact north of \sgA\ on the southern boundary of nebula SmR-3. The region plotted is delineated by box (a) in Figure~1. (a) Line-only emission after subtracting continuum using the NICMOS/F190 image as described in \citet{2011MNRAS.417..114D} and extracted from the HST/NICMOS \pA\ Survey of the Galactic Centre \citep{2010MNRAS.402..895W}. The jet axis defined at a smaller scale near the CND is extrapolated as the dotted red line. The 10\arcsec$\times$6\arcsec\ and 20\arcsec$\times$10\arcsec\ (magnified as Figure~\ref{fig:nshock} top) boxes delineate where we summed \pA\ flux. (b) NICMOS F187N filter image; the nearest WN star lies just off the top right of this figure. (c) JVLA 5.5 GHz continuum (from Z16) matching (a) well.}
    \label{fig:Palpha}
\end{figure*}

T18 identified other structures whose kinematics deviate from the clockwise-rotating CND. Here, we mention only two, first redshifted ``Anomaly A" that ``counter-rotates" \citep{2012A&A...539A..29M} by $\sim50$\kms\ on the near side to us in the Western Arc (see Figure~\ref{fig:CND}(d)).
It is clearest at 67--117.5\kms\ in CS, C$^{34}$S, and SiO (which tracks fast shocks), but is faint in H$^{13}$CO$^+$.
In SiO $J=2-1$ its filaments span $\sim100$\kms, broader than other features around the CND (T18 Figure~8(d)), that extend for $\sim1$\arcmin\ from its northern edge.
T18 estimated an LTE ionized mass of $(1200-6500) (T_\mathrm{ex}/100$ K) \Msun\ to optical depth 1.5.
Therefore, like Anomaly C, it is shock excited but here stronger at $\sim50$\kms\ and potentially several times more massive.
T18 suggested that this gas is infalling to \sgA, but our simulations described in Section \ref{sec:sims} show a broad pattern of jet induced outflow from the CND that generates filaments with kinematics like Anomaly A.

\subsubsection{``Fork" at Anomaly D of T18}
The northern jet path also projects onto a \pA\ and radio continuum ``fork" (Figure~\ref{fig:PAnearGC} top) before reaching Anomaly C.
Our SOAR/tspec4.1 near-IR spectra along this structure (Section 3.4.2) show H$_2$ S- and Q-branches, Br hydrogen, and weak He I emissions.
T18 Figure~1(f) shows CS and \citet{2017ApJ...842...94T} Figure~4(d) shows H42$\alpha$ emissions here on the ionized southern end of Anomaly D.
It is prominent between -72.5 and -32.5\kms\ versus $\sim10$\kms\ in the adjacent CND, but is too faint in other millimeter wavelength lines for T18 to establish its physical properties other than that its lack of CH$_3$OH places it near \sgA.
We therefore do not discuss it further.

\subsection{Jet Influence beyond 2 pc North of the GC?}\label{sec:Njet}

\subsubsection{Prior work}

Beyond the ALMA data cube, $6\times12$ pc bipolar ellipsoids anchored on the CND \citep{2003ANS...324..167M} are prominent in Ar-Ca + Blue-Ca and Ca XIX bands \citep[photon energies 3.27--3.73, 4.07--4.5, and 3.78--3.99 keV, respectively,][P15 hereafter]{2015MNRAS.453..172P}.
Each contains 1--3 $M_\odot$ of hot gas with thermal energy $\sim10^{50}$ erg.
Their shocked edges brighten at 4\arcmin\ from the GC, which suggests to those authors an explosive not stellar-wind origin.
No compact features to the north in the Chandra image (e.g.,~Figures~10--12 of P15) align with the jet axis, but the XMM-Newton bands show (see Figure~12 of P15) that the north lobe at 5\farcm4 (13 pc on sky) has a bright edge across a plausible jet span.
Closer in, Figure~\ref{fig:ChandraPA} shows that the north lobe is embedded within a broad cone ($\sim75$\arcdeg\ opening angle) of radial dust plumes and \pA\ \citep[Figure~2]{2010MNRAS.402..895W} but no CS($J=2-1$) (Figure~7 of  \citet{2016ApJ...831...72H}).
Position-velocity diagram (PVD hereafter) in Figure~\ref{fig:Nobfig} shows that
CS with kinematics like those of Anomaly C span 110 \kms\ and extend north for 1\farcm8 to almost  $b=0$\arcdeg. (We explain shortly how SOAR telescope near-infrared spectra registered the \pA\ HST image to this PVD of CS.)

\subsubsection{New insights on nebula SmR-3}

\begin{figure}
    \centering
    \epsscale{1.18}
    \plotone{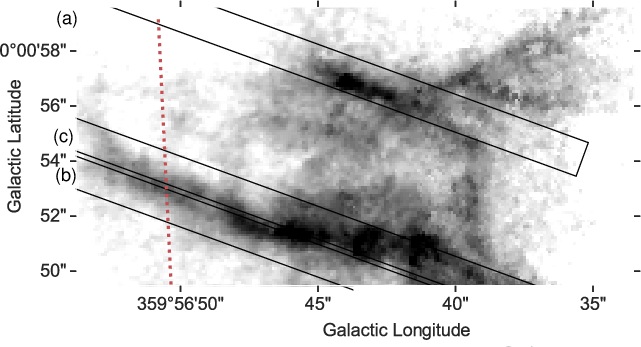}
    \plotone{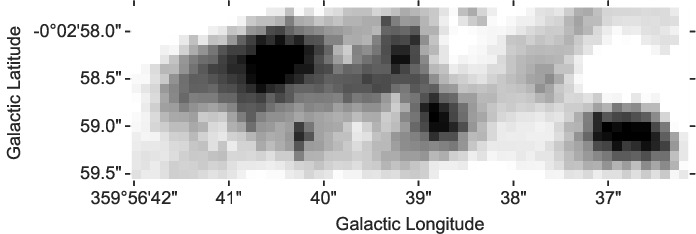}
    \caption{The proposed jet/ISM complexes in \pA\ south (bottom, from box in Figure~\ref{fig:PAnearGC}), and north (top, showing the three SOAR/tspec4.1 slits positioned along 12\arcdeg\ PA over the region delineated by the larger box in Figure~\ref{fig:Palpha}).}
    \label{fig:nshock}
\end{figure}

Nebula SmR-3 is 3\farcm6 from the GC (8.6 pc on sky, Figure~\ref{fig:Palpha}). Z16 derived its 5.5 GHz flux density of $0.41\pm0.06$ Jy, assuming thermal emission.\footnote{We used the 5.5 GHz image of Z16 from Dr.~Zhao's website because the JVLA archive provided only \textit{uv} tracks that would require specialized processing beyond the scope of this paper to optimize dynamic range of the derived image.}
Across a third of its southern boundary (Figure~\ref{fig:nshock} top) is a bright wedge perpendicular to the projected jet axis that would span at least 3\arcdeg\ as seen from \sgA.

Within the boxes in Figure~\ref{fig:Palpha}(a) we subtracted the underlying diffuse emission then summed the \pA\ flux.
The bigger box has $\sim3.9\times10^{30}$ \ergs, while the smaller has $1.7\times10^{30}$ \ergs. These straddle the luminosity of \pA\ at the southern shock (Figure~\ref{fig:nshock} bottom).
Likewise panel (c) yields 120 and 15 mJy, respectively, i.e.\ the V-feature in radio contains a smaller fraction of the southern boundary flux.
Just south of SmR-3  we find an average surface brightness over $3\times3$ arcsec$^2$ of 0.12~mJy per $1.6\times0.6$ arcsec$^2$ beam versus 0.44~mJy beam$^{-1}$ at its southern boundary.
Converting these optically thin values to volume emissivity requires the length along our sightline.
Figure \ref{fig:Palpha}(c) shows that most of the radio emission below the southern boundary of SmR-3 lies within filament NWstr-3 of Z16, originating near \sgA\ and not distributed throughout the GC region.
It is barely resolved, $\sim1\farcs5$ across, so we assume this depth along our sightline.
The on/off flux ratio here is less than four so is not a J-shock front, but instead might be ionized by the WR and other hot/windy stars nearby, or be a C-shock (ions too weak to dissociate the ambient, magnetized ISM).

\begin{figure}
    \centering
    \includegraphics[scale=0.6]{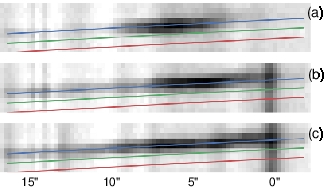}
    \caption{Br~$\gamma$ emission-line profiles along the slits in Figure~\ref{fig:nshock} with isovels at -75, 0, +75 \kms\ relative to the LSR velocity shown in blue, green, and red, respectively. The vertical bands are stars and residual OH$^+$ sky lines.}
    \label{fig:brgamma}
\end{figure}

To constrain its location and excitation, on UT 2021 May 29 $\lambda\lambda$1--2.4 \micron\ SOAR/tspec4.1 \citep{2014SPIE.9147E..2HS,2020SPIE11447E..6LH} echelle spectra were obtained along the three slits shown in Figure~\ref{fig:nshock} at spectral resolution $R\sim3500$ with a 1\farcs16 wide slit in 0\farcs75 FWHM seeing.
Several emission lines were mapped including the He~I triplet from hot stars in the GC; after accounting for the instrumental profile of $\ga80$\kms, all were narrower than 6\kms\
without detectable broad wings. The nebula is likely on the near side of the GC because the Pa~$\beta$/Br~$\gamma$ flux ratio of $\sim2.5$ that we measure indicates a K-band extinction characteristic of that region \citep{2011ApJ...737...73F}.

These spectra, along with those obtained along Anomaly D, and their relation to the radio continuum emission will be discussed in a subsequent paper, but Figure~\ref{fig:brgamma} shows Br~$\gamma$ emission-line profiles along the slits across SmR-3 with wavelengths simultaneously calibrated from OH$^+$ lines and the velocity scale corrected to the LSR.
All velocity centroids are at -62$\pm5$\kms, placing SmR-3 25\kms\ from the centroid of the blueshifted side of the split CS line profile in Figure~\ref{fig:Nobfig}.
Section \ref{sec:discussN} discusses the implications of this velocity.

\section{Simulations}\label{sec:sims}

Our numerical simulations examine whether the observations described above could trace the effects of an AGN jet. The main difficulty in modeling an MW jet and its ISM interactions is that we know neither the jet properties nor the ISM conditions when jet activity starts. Prior jet events are also likely, in which case we also do not know the duty cycle. While a study of the full parameter space of the jet and ISM properties is beyond the scope of this work, we can nevertheless create, without excessive tuning, models that match the observed gas-dynamical phenomena and morphological structures. 

We performed a series of 3D relativistic hydrodynamical simulations of an AGN jet launched from the MW central black hole that interacts with the surrounding ISM on two separate spatial scales: (1) the central kiloparsec-scale volume of the MW disk to compare its morphology with the MeerKAT 1.274 GHz and the 1.5--2.6 keV XMM-Newton X-ray data sets; (2) the CND-scale GC to focus mostly on the formation of spatio-kinematical molecular structures resembling Anomaly C.

We constructed initial conditions that satisfy the observational constraints described in Sections 2 and 3 and that represent the gravitational potential and gas distribution of the central MW. The Appendix~\ref{s:simulations} details how the simulations are set up and conducted. The jet power is fixed at $\Pjet=10^{41}$ \ergs.
Simulation results including movies are available from \url{https://www2.ccs.tsukuba.ac.jp/Astro/Members/ayw/research/mwagn}.

\subsection{Results from the Kiloparsec-scale Simulations}\label{sec:kpcscale}

\begin{figure}
    \centering
    \includegraphics[width=0.8\linewidth]{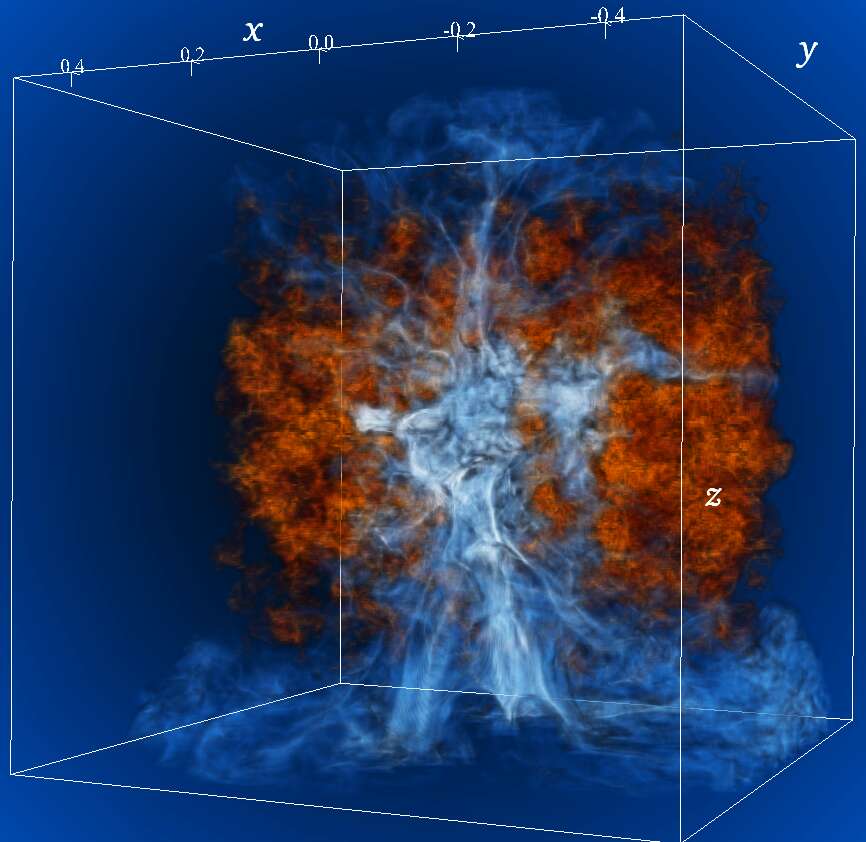}\\[-0.1mm]
    \includegraphics[width=0.8\linewidth]{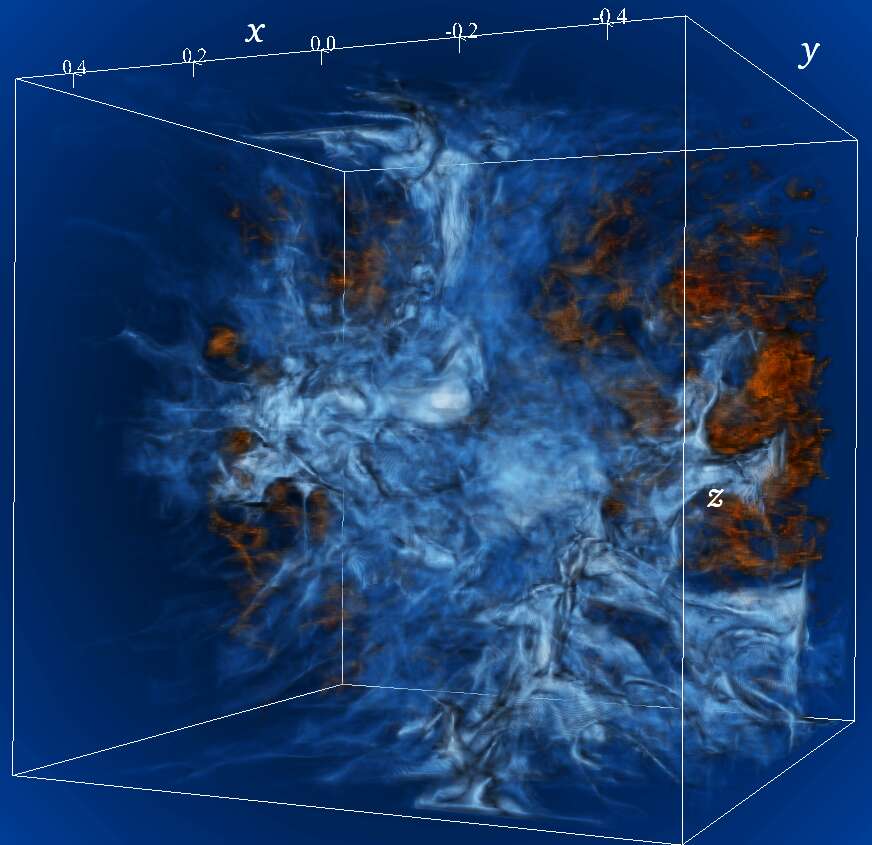}
    \caption{Volume renders of a GC-scale simulation to show how extensively the jet plasma percolates through the clumpy ISM to disperse the clouds. Each is a composite of the jet tracer in bluish-white and the density of dispersed clouds in orange. Top snapshot at $t\approx0.5$ Myr, bottom at $t\approx3$ Myr. Axis labels are in kpc.}
    \label{f:kpc-vr1}
\end{figure}

Figure \ref{f:kpc-vr1} is a volume rendering of two snapshots, at  $t\approx0.5$~Myr and $t\approx3$~Myr, of the simulation to display the expulsion of the ISM clouds (density in orange) and the spatial extent of the jet plasma by using the jet tracer variable (bluish-white). 

\begin{figure*} 
    \centering
    \includegraphics[width=0.85\textwidth]{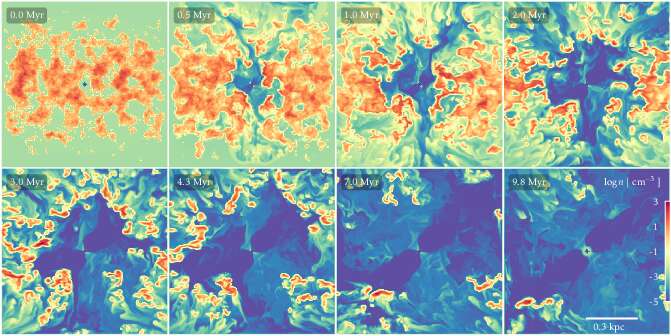}
    \caption{Midplane slices at $y=0$ of the evolving density field. In this color map, jet plasma is bluish, hot ISM pale green, and warm and cold clouds are orange--red. Jet plasma can displace dense gas in the central kpc$^3$, consistent with current day observations of the HI and molecular gas density distributions in the MW.}
    \label{f:kpc-ev1}
\end{figure*}

\begin{figure*}
    \centering
    \includegraphics[width=0.85\textwidth]{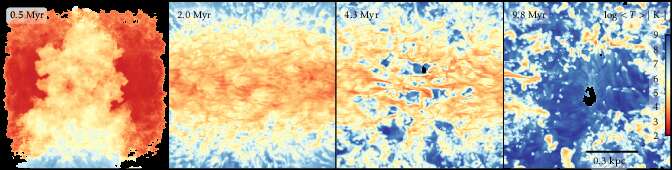}
    \caption{Evolution of the mean gas temperature projected along the $x$-axis (sightline toward \sgA). In this color map, warm clouds are orange--red, while hotter gas including gas ablated from clouds is yellow and blue. Despite the strong dispersal of clouds by the jet, the cloud cores remain cool due to their efficient radiative cooling.}
    \label{f:kpc-temp}
\end{figure*}

The snapshots in Figure~\ref{f:kpc-ev1} show the jet propagating through the height of the MW disk and strongly interacting with ISM clouds. While the outer layers of the clouds are slowly ablated by the shearing jet streams, layers within are soon compressed by radiative shocks and begin to contract into long-lived filaments that fragment due to hydrodynamical and thermal instabilities. Clumps and filaments are slowly carried outward at $\sim100$\kms{}. Snapshots in Figure~\ref{f:kpc-temp} of the mean gas temperature (excluding any jet plasma) along our sightline ($x$-axis) at a fairly late time $t=9.8$ Myr show that the clouds, although dispersed by strong interactions with the jet, radiate efficiently to remain cool.

As the main jet stream interacts with clouds, it splits into multiple secondary streams that percolate through the entire volume to disperse and gradually evacuate clouds from the central region around the main jet path. Even after the jet head has propagated out of the simulation box, strong interactions persist as small clouds in the jet path split and divert jet streams. Throughout the simulation, clouds experience strong ram and thermal pressures from the jet \citep{Wagner2012-cx}.

To  find a jet powerful enough to not be entirely frustrated at the end of a simulation but weak enough to interact with a substantial gas volume, we tested ranges of jet powers, $\Pjet$, and their associated over-pressures with respect to the ambient medium, bulk Lorentz factors, $\Gamma$, and ratios of jet rest mass energy density to jet pressure, $\chi$. Jet powers of $10^{42}$ \ergs{} $> \Pjet > 10^{40}$ \ergs{} are satisfactory; here, we  present only the results using the fiducial parameters $\Pjet = 10^{41}$ \ergs{}, $\Gamma = 1.1$, $\chi = 50$, and radius $\rjet = 12$ pc. At its base the jet is over-pressured thirteen-fold, and its density relative to the adjacent ISM hot-phase gas is $\sim$0.0025.

We also tested a range of mean column densities for the clouds. We settled on $\Nav\approx1.5\times10^{21}\ppcs$ because smaller values excessively mix jet plasma, hot-phase gas, and clouds, while larger columns produce excessive flow variations for different cloud initializations, i.e., clouds are too large and too few to randomize the jet flow sufficiently. 
We further discuss the role of $\Nav$ in Section \ref{s:sim-disc}.
Simulations with this column match the synthetic radio morphology of the jet to the MeerKAT 1.274 GHz data. 

\begin{figure*}
    \centering
    \includegraphics[width=0.85\textwidth]{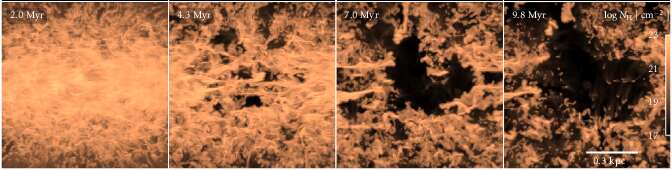}
    \caption{Evolution of the gas column density along the $x$-axis (sightline toward \sgA) when $t>2$ Myr. After $\sim5$ Myr much of the gas in the central region is pushed away by the jet so that after 10 Myr much has left the inner kpc$^3$.}
    \label{f:kpc-Nh}   
\end{figure*}

\begin{figure*}
    \centering
    \includegraphics[width=0.9\textwidth]{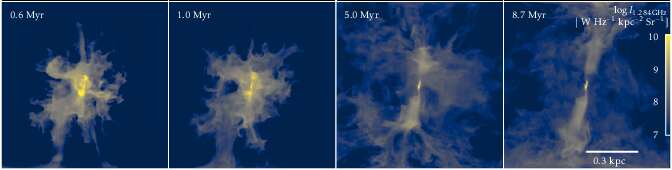}
    \caption{Evolution of the radio surface brightness at 1.28~GHz after free-free absorption, as seen along the $x$-axis (sightline toward \sgA). The ray-tracing is performed as in \citet{Bicknell2018-ag} (see their Appendixes). Although confined, the main jet stream splits into secondary streams that flood through the domain to keep the ISM hot. The main jet stream clarifies after 8 Myr when the jet, which is over-pressured compared to ambient gas, has cleared away many clouds in its path so can propagate more freely to the edge of the simulation domain. Figure~\ref{fig:42ergJet} compares this simulation to the MeerKAT radio image.}
    \label{f:kpc-radio}   
\end{figure*}

\begin{figure*}
    \centering
    \includegraphics[width=0.8\textwidth]{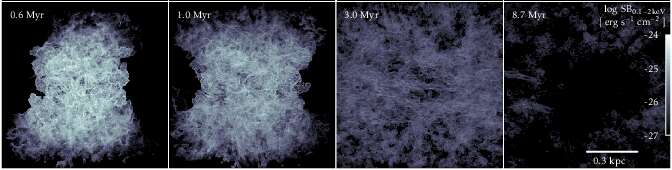} \\[-3.7mm]
    \includegraphics[width=0.8\textwidth]{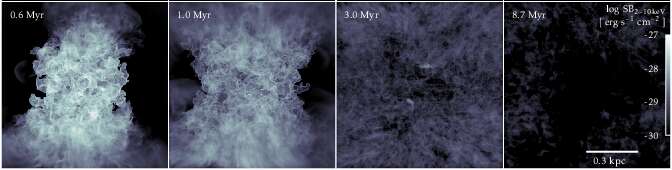}
    \caption{The evolving soft X-ray (0.1--2 keV without Galactic absorption, top row) and hard X-ray (2--10 keV, bottom row) surface brightnesses as ray-traced along the $x$-axis (sightline toward \sgA) at four different times. The ISM heats quickly at bow shocks of the jet-driven bubbles and at shocks generated by the jet streams. The cooling layer of radiative shocks propagating into the clouds is clear in the soft X-ray images, while the hard X-rays highlight regions of strong interactions and sometimes show filaments associated with ablated hot gas. The surface brightness fades rapidly after $\sim5$ Myr once central X-ray cavities form.}
    \label{f:kpc-xs4}   
\end{figure*}

\begin{figure*}
    \centering
    \includegraphics[width=0.85\textwidth]{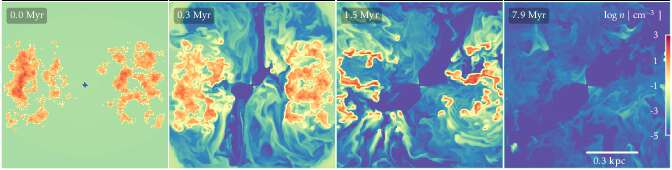}
    \caption{Midplane slices at $y=0$ of the evolving density field of the simulation with $\rmmm=0.3$. Initially the central kiloparsec region is much more gas-poor compared to the $\rmmm=0$ case shown in Figure~\ref{f:kpc-ev1} because it starts out partially evacuated by prior AGN activity. In this color map, the jet plasma is bluish, the hot ISM pale green, and the warm and cold clouds are orange--red.}
    \label{f:kpc-ev2}   
\end{figure*}

Figure~\ref{f:kpc-Nh} shows how the gas column density along the $x$-axis evolves after 2 Myr. Gas in the central region evacuates over a few Myr. After $\sim5$ Myr much has been pushed away by the jet.  At $t=9.8$ Myr, the column density snapshot and Figure~\ref{f:kpc-temp} map of projected mean temperature show a 250 pc radius, spherical shell of accumulated clouds. Note that the central region continues to evacuate long after the jet has traversed the height of the MW disk because secondary albeit weakening jet streams and lateral pressure gradients around clouds both persist.  Our simulations thus support the idea that a jet operating over the most recent few Myr is responsible for the present-day deficit of molecular and neutral gas in the inner kiloparsec of the MW.

By $\sim10$ Myr, much gas has left the central kiloparsec but dense clumps remain as the jet loses its mechanical advantage after breaking out to flow more freely through the inner kiloparsec \citep{Wagner2011-rh}. This simulation shows that $\Pjet = 10^{41}$ \ergs\ is near the minimum power required to clear the central kiloparsec of the MW of gas.

Figure~\ref{f:kpc-radio} shows snapshots of the synthetic radio surface brightness at 1.28~GHz of the simulation as seen from Earth. 
 
The radio emissivity and free-free absorption was computed as in \citet[see their Appendices]{Bicknell2018-ag}. The radio plasma quickly fills the entire volume to make the inner kiloparsec very radio bright. Rather than appearing as a highly collimated beam, the radio jet is broadened by its over-pressure and is split into secondary streams by clouds in its path.

Figure~\ref{f:kpc-xs4} shows soft (0.1--2 keV) and hard X-ray (2--10 keV) surface brightnesses obtained directly from the precomputed cooling function of MAPPINGS V \citep{Sutherland2017-ep}, non-ionization-equilibrium cooling calculations that are also used in the simulations. The MAPPINGS V cooling rates per unit density squared at the cell temperature for these two energy bands are multiplied by the cell density squared and integrated along lines of sight. The MAPPINGS V calculations assumed a thermal plasma and the X-ray emission in the two energy bands is predominantly free-free emission.
The Figure shows that the ISM heats quickly by bow shocks of the jet and collisions, and by shear forces between jet flows and clouds. By 0.5 Myr the bow shock has propagated to the domain boundary. 

While the main jet stream remains confined in the inner few hundred parsecs, secondary-jet streams flood the entire domain to keep the X-ray gas hot. The cooling layer of radiative shocks that are propagating into the clouds is especially clear in the soft X-ray images. The hard X-ray surface brightness is associated more with ablated gas that has undergone strong mixing with the hot-phase gas and jet plasma. It is more uniform later at $t=3$ Myr except at bright hot spots where direct head-on interactions with the jet drives particularly strong shocks into dense clouds.

After the jet has pushed beyond 500 pc radius, it substantially reduces the filling factor of clouds to make the radio surface brightness more uniform. The jet morphology remains laterally extended, not as narrow as one might expect, because jet plasma still spreads because of its interaction with remnant cloudlets.  At $t>8$ Myr the central region becomes an X-ray cavity because the jet has pushed away most of the thermal gas. The morphological differences of the X-ray emitting gas persist, with the hard X-ray maps showing filamentary ablated gas pointing toward the Galactic plane. These head-tail structures form when the jet backflow ablates clouds. 

Outbursts may be cyclical. We therefore repeat the simulations described above, changing only the initial central gas distribution parameter of the \citet{McMillan2017-kl} profile for the HI disk to $\rmmm=0.3$ to partially fill the central hole. This represents the gas distribution once the MW disk gas has somewhat resettled following a prior outburst.
Figure~\ref{f:kpc-ev2} shows the evolution of the jet and clouds along the midplane slice $y=0$. The evolution is similar to that of the simulation with $\rmmm=0.3$, but now the jet plasma vents through the Galaxy much more quickly so that most of the plasma has escaped by 1.5 Myr. The tilt of the jet ensures that jet--cloud interactions persist beyond several Myr, but are fewer and weaker than those seen when $\rmmm=0$. The central region has little dense gas by 5 Myr. 

Clouds carried away by the jet plasma either form long filaments later on (e.g., Figure~\ref{f:kpc-ev2} 1.5 Myr panel) or, if the jet stream and cloud collide strongly, flatten perpendicular to the stream and trail ablated gas \citep{MacLow-mk}. An example of the latter can be seen in Figure~\ref{f:kpc-ev1} in the top-right quadrants of the snapshots for $t=3.0$ Myr and $t=4.3$ Myr, where the jet hits clouds. Note how the cloud resembles the southern boundary of SmR-3 seen in Pa$\alpha$ and JVLA 5.5 GHz (Figure~\ref{fig:nshock}).

\begin{figure*}
    \centering
   \includegraphics[width=0.83\textwidth]{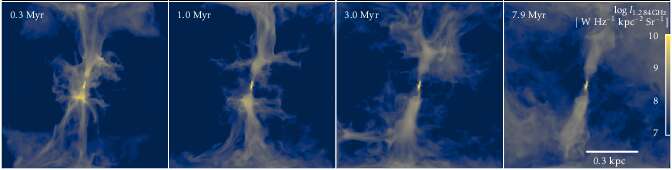}
    \caption{Evolution of the radio surface brightness at 1.28~GHz as seen parallel to the $x$-axis, as in Figure~\ref{f:kpc-radio}, but for the simulation where $\rmmm=0.3$.}
    \label{f:kpc-radio-2}   
\end{figure*}

\begin{figure*}
    \centering
    \includegraphics[width=0.82\textwidth]{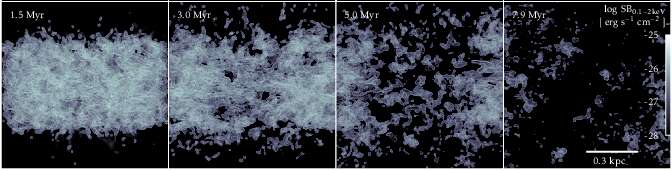}\\[-3.7mm]
    \includegraphics[width=0.82\textwidth]{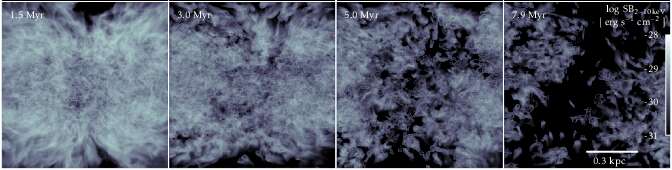}
    \caption{Evolution of the soft X-ray (0.1--2 keV without Galactic absorption, top) and of the hard X-ray (2--10 keV, bottom) surface brightnesses after 1.5 Myr for the simulation where $\rmmm=0.3$. The view parallels the $x$-axis (sightline toward \sgA).}
    \label{f:kpc-xs-r-2}   
\end{figure*}

Figure~\ref{f:kpc-radio-2} shows how the radio surface brightness evolves in this simulation. In the first snapshot at $t=0.3$ Myr the main jet stream interacts strongly with clouds in the central region to disperse jet plasma in all directions, but mainly into two or three longer, coherent streams that are propagating into the halo. Splitting of the main stream is most pronounced early on. At $t=3$ Myr we see an asymmetric jet morphology: the northern jet still strongly interacts with a cloud while the southern jet streams into the halo largely unimpeded. Later, at $t=8$ Myr, both jets are propagating freely into a halo filled with jet plasma. The jets seem to become turbulent 300-400 pc from the central BH, possibly due to their deceleration that may be somewhat amplified by effects of our computational boundary.

Most of the soft X-ray emitting gas shown in the upper panels of Figure~\ref{f:kpc-xs-r-2} is confined to the inner disk, while the hard X-ray emitting gas shown in the lower panels includes swept-up material that has been carried into the upper region of the disk and the halo. The cavities in the disk blown by the jet lead to patchy surface brightness in the X-ray maps, particularly clear in the soft X-rays later on ($t>3$ Myr). Small X-ray cavities in the disk appear to have typical radii of $\sim50$ pc at $t=3$ Myr, 100 pc at $t=5$ Myr and 100 pc at $t=8$ Myr. Typical sizes of X-ray bright structures are defined mainly by the clouds, $\sim50$ to 100 pc, that decrease slightly over time as the clouds are ablated, compressed, and fragmented.

\subsection{Results From the CND-scale Simulations}

The two simulations with different CND initial conditions reveal different ways to form Anomaly C. In both, the jet hits the front and back edges of the CND directly to start the strongest interactions.

\subsubsection{Smooth CND}\label{s:sim-smooth-cnd}

\begin{figure*}
    \centering
    \includegraphics[width=0.8\textwidth]{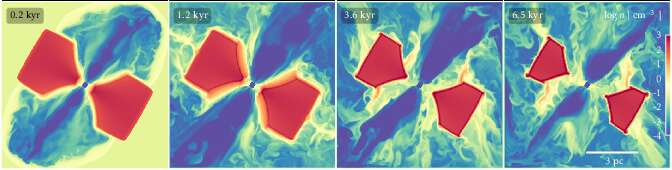}
    \caption{Midplane slices at $y=0$ of evolution of the density field of the simulation with the smooth CND. In this color map, the jet plasma is bluish, the hot ISM is pale yellow--green, and the warm and cold clouds are orange--red.}
    \label{f:pc-ash}   
\end{figure*}

\begin{figure}
    \centering
    \includegraphics[width=0.9\linewidth]{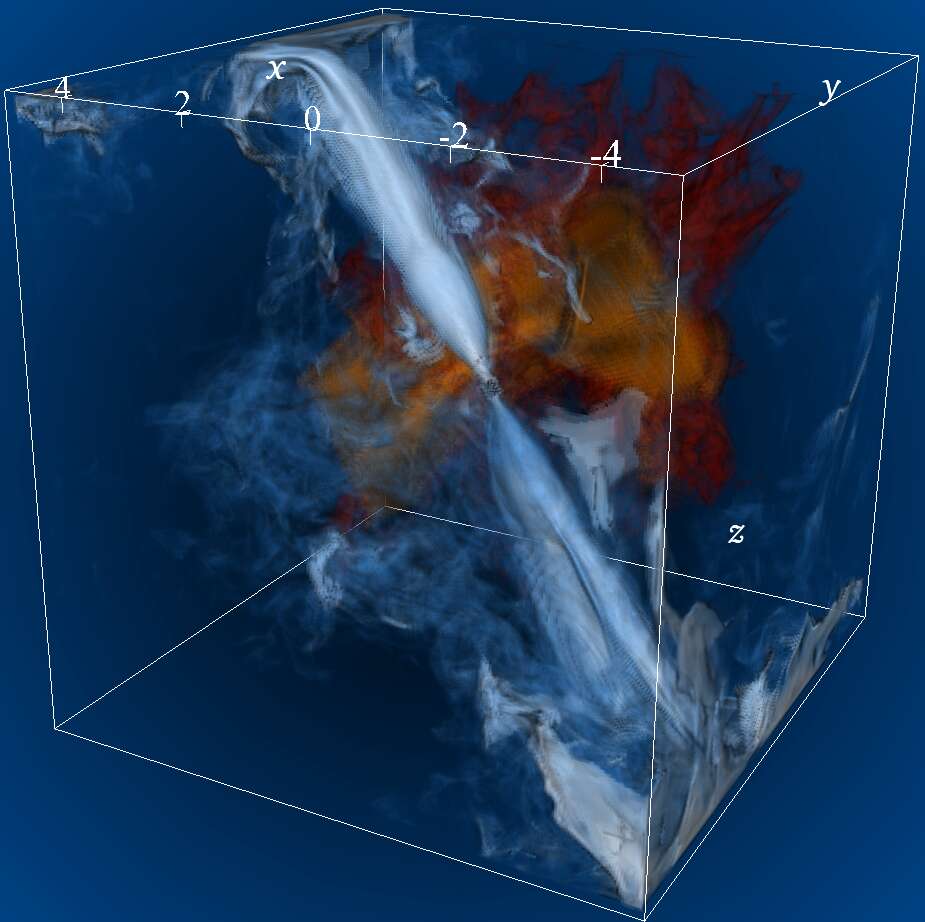}
    \caption{Volume render of the smooth CND simulation at $t=6.0$ kyr with axis labels in pc. The jet tracer is bluish-white and the CND gas density is orange--red.}
    \label{f:pc-vr2}   
\end{figure}

For the smooth CND, no jet plasma may enter, so the CND remains largely intact throughout the simulation. Figure~\ref{f:pc-ash} shows the evolution of the density of the jet and CND along a midplane slice through $y=0$. Figure \ref{f:pc-vr2} shows the volume render of the jet tracer and CND gas density for the $t=6$ kyr snapshot. Jet plasma engulfs the CND and drives a strong radiative shock within that quickly stalls to form a dense protective layer, except where the jet hits the CND directly in ``splashes."
There, shear instabilities grow at the surface that ponderously ablates gas and entrains as filaments into the jet while it is active.

The filaments stretch along the edge of the jet stream. Their density is fairly low ($0.1\ppcc$ to $1\ppcc$) and their velocities are high (100-1000\kms{}), although it is unclear to what extent numerical diffusion is responsible for the heating that facilitates entrainment and acceleration. Regarding morphology, while it is inviting to interpret the formation of filaments as a possible origin of Anomaly C, their total mass is only 100 \Msun. Perhaps we have simulated the interaction of the jet too briefly. But, more likely, our highly idealized setup to explore direct ablation and entrainment of filamentary gas from the smooth surface of the CND does not permit thermal instabilities that would further mass-load the jet entrainment.

After $t=3.6$ kyr until the run ends at 6 kyr the CND continues to shrink as gas ablates and accretes onto the BH, and the jet continues to entrain the ablated gas from the surface of the CND. 

\begin{figure*}
    \centering
    \includegraphics[width=0.82\textwidth]{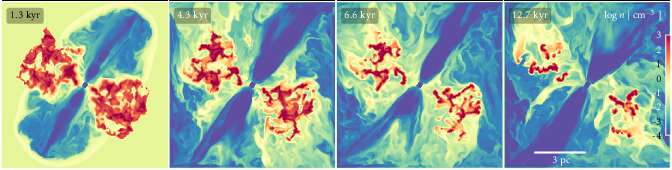}\\[-3.1mm]
    \includegraphics[width=0.82\textwidth]{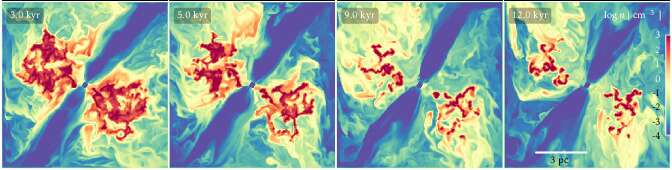}
    \caption{Midplane slices at $y=0$ of the evolving density field of two simulations with a clumpy CND. The slightly different initial density distribution in the bottom row increases the head-on collisions of the jet with clumps in its path. It shows stronger head-tail structures of clumps that are lifted from the CND. In the color map, the jet plasma is bluish, the hot ISM is pale yellow--green, and the warm and cold clouds are orange--red.}
    \label{f:pc-a6hte}   
\end{figure*}

\begin{figure}
    \centering
    \includegraphics[width=0.85\linewidth]{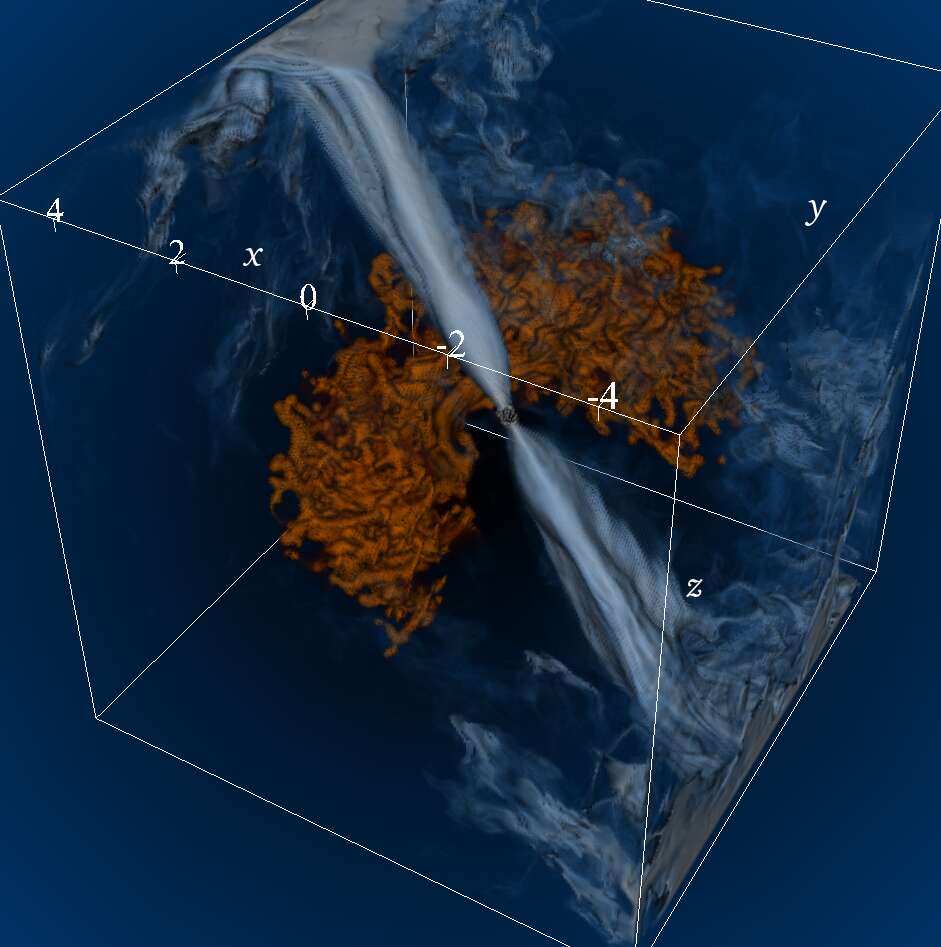}
    \caption{Volume render of the clumpy CND simulation at $t=7.0$ kyr with axis labels in pc. The jet tracer is bluish-white and the CND gas density is orange--red.}
    \label{f:pc-vr3}   
\end{figure}

\subsubsection{Clumpy CND}

Figure~\ref{f:pc-a6hte} shows simulations with an initially clumpy CND made turbulent by jet interactions. They evolve very differently from the smooth case just discussed because bow shocks and jet plasma can propagate within the CND. Midplane slices in the top row of Figure~\ref{f:pc-a6hte} and the volume render of the jet tracer variable and CND gas density (Figure~\ref{f:pc-vr3}) are shown at $t=7$ kyr. As the jet plasma percolates into the CND through lower-density channels, the CND gas disperses and shocks. Its inhomogeneities seed its fragmentation into dense clumps and its contraction is induced by thermal instabilities. It gradually flattens while becoming more porous. As it begins to contract and ablate, we see two developments not apparent in the smooth CND simulations that may be associated with Anomaly C.

First, the gas contracts into elongations that tend to be either parallel or perpendicular to jet streams that are percolating through the CND. Vertical dense plumes appear while more diffuse gas is removed continually from the CND, taking away substantial angular momentum and contracting the CND toward the midplane of its rotation. The plumes contain more gas than the ablated filaments (a few hundred~\Msun{}) and also have much lower velocities because they are not entrained along the jet. They nonetheless result directly from jet--gas interactions: radiative shocks driven by the percolating jet streams trigger their runaway collapse. Their life-span is uncertain, but may be fairly long because their high density minimizes hydrodynamical ablation.

Second, gas clumps are slowly lifting off the edges of the CND. They form cometary head--tail structures that persist due to radiative cooling and continuous compression by shocks. This is more pronounced in the simulation shown in the bottom row of Figure~\ref{f:pc-a6hte}, which has a slightly different instance of initial densities so that the filling factor of clouds ahead of the jet is slightly higher to increase head-on collisions with the jet. Such clumps accelerate over a few kyr before dissipating.

Anomaly A may form from CND filaments that are contracting radiatively as they are compressed by jet plasma and gas turbulence incited by the jet. Gas that has been strongly impacted by the jet can diverge strongly from the mean rotation curve of the CND, as seen on galaxy scales in the molecular disk of radio Seyfert galaxy IC 5063 \citep{Mukherjee2018-nd}.

\subsubsection{Relaxation After Jet-CND Interactions}\label{s:sim-relax}

\begin{figure*}
    \centering
    \includegraphics[width=0.66\textwidth]{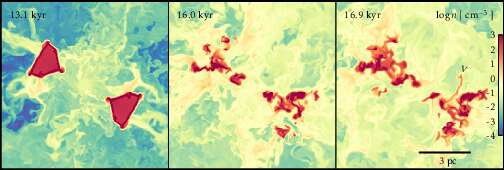}
    \caption{Midplane slices at $y=0$ of the density field of the three simulations presented in Figures~\ref{f:pc-ash} and \ref{f:pc-a6hte}, after turning off the jet so that the turbulent CND and ambient gas relax for a few kyr. After the jet is off, much of the warm and hot gas expands. Left: simulation with the smooth CND torus (see Figure~\ref{f:pc-ash}); Middle: simulation with a clumpy CND torus (see top row Figure~\ref{f:pc-a6hte}); Right: simulation with an alternate clumpy CND torus (see bottom row Figure~\ref{f:pc-a6hte}). The arrowhead at right points to the narrow filament noted in Section \ref{s:sim-relax} that resembles Anomaly C.}
    \label{f:pc-relax}
\end{figure*}

To explore relaxation of the ISM, we followed evolution after switching off the jet at 3.0, 2.8, and 4.2 kyr of jet--CND interactions in the simulation with the smooth and the two simulations with the clumpy CND, respectively.  Figure \ref{f:pc-relax} shows snapshots of the last time step of the relaxation.

For both the smooth and clumpy CND, the jet plasma, by virtue of its high internal pressure and radial momentum gained through interactions, continues to expand outward adiabatically. Although turbulent motions persist, turbulent and thermal pressures around the CND gradually decrease, and the dense CND gas begins to mix with the ambient ISM. The filaments cease to be compressed by strong shear from a continuous jet stream, so that most of the outflowing filaments and even clumps begin to expand rapidly with the jet plasma. 

However, some filaments recondense or form from the highly mixed gas. A prominent example is the long feature denoted by the arrowhead at $(x,z) \approx (2.5,0)$ pc in the right panel of Figure~\ref{f:pc-relax}. The density of this filament is intermediate to the that of the sharp directly ablated material seen in the smooth CND simulation and that of the thermally collapsing features seen inside the CND in the clumpy CND simulations.

\subsubsection{Position--Velocity Diagrams}

PVDs clarify more aspects of the jet--CND interactions by giving insight into the resulting velocity structures. They were created by mass-weighting 2D histograms in the $y$ coordinate (the dimension we see across the MW disk) and $v_x$ (the velocity component along our sightline) of all simulation cells. Note that the simulated PVD and associated column density maps cannot be compared to, e.g., CO data cubes because the observed gas is often optically thick.

For the simulations with a smooth CND, no features in the PVD were associated with filaments created in the ``splash" region where the jet is ablating the CND because the ablated gas densities were too low to produce features discernible from the unperturbed CND. 

\begin{figure*}
    \centering
    \includegraphics[height=0.27\textwidth]{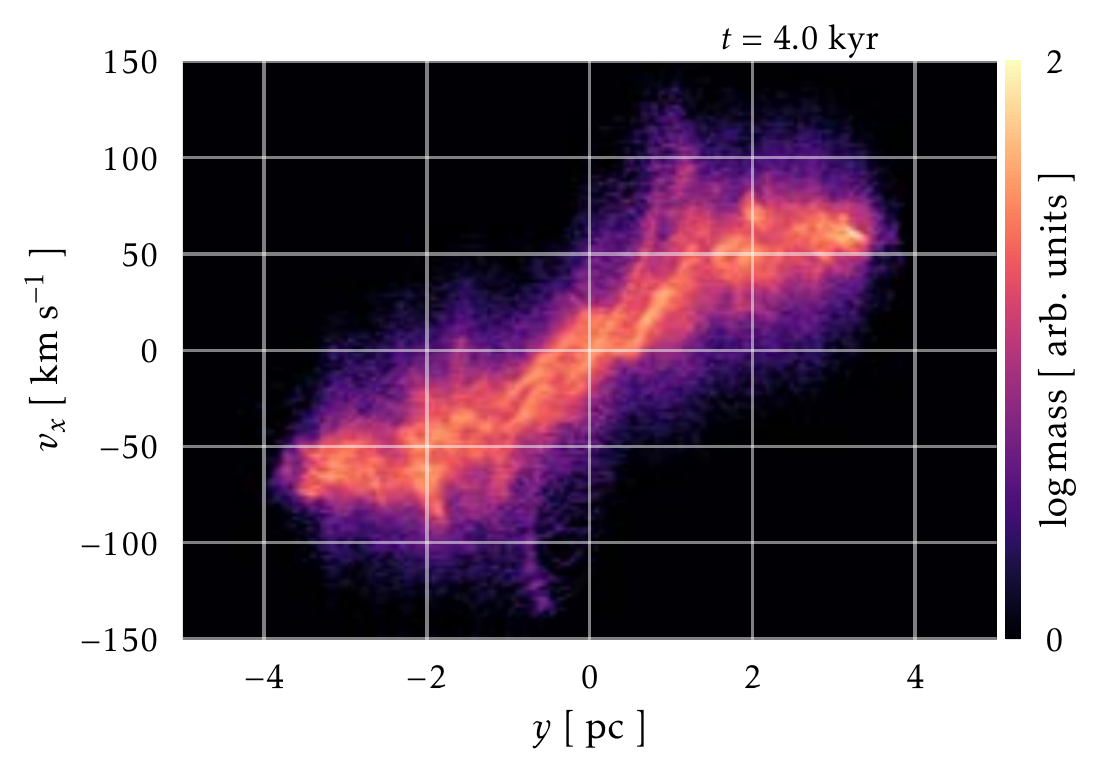}
    \includegraphics[height=0.27\textwidth]{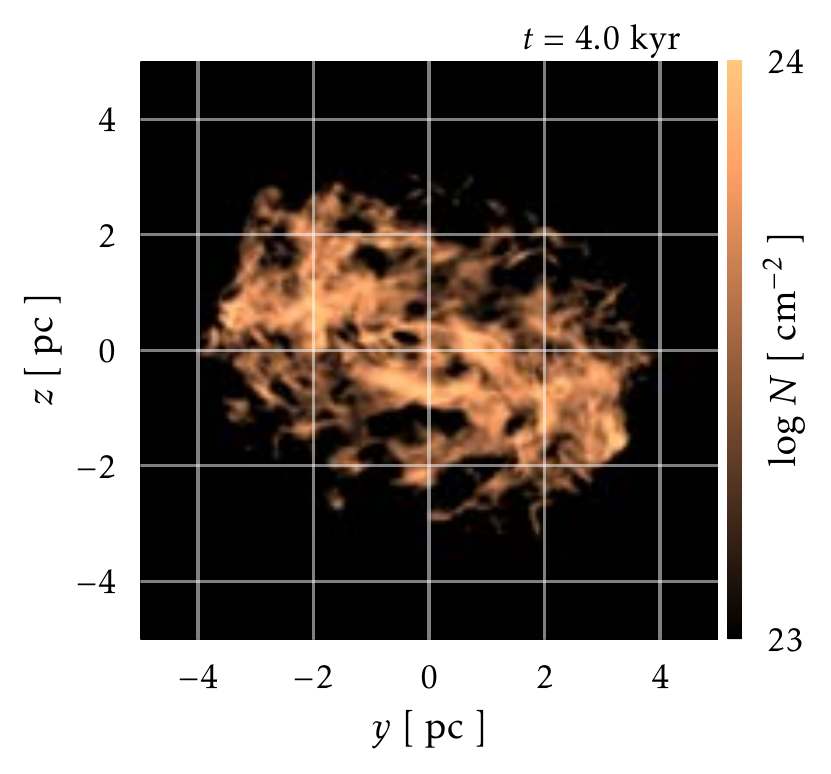}
    \includegraphics[height=0.27\textwidth]{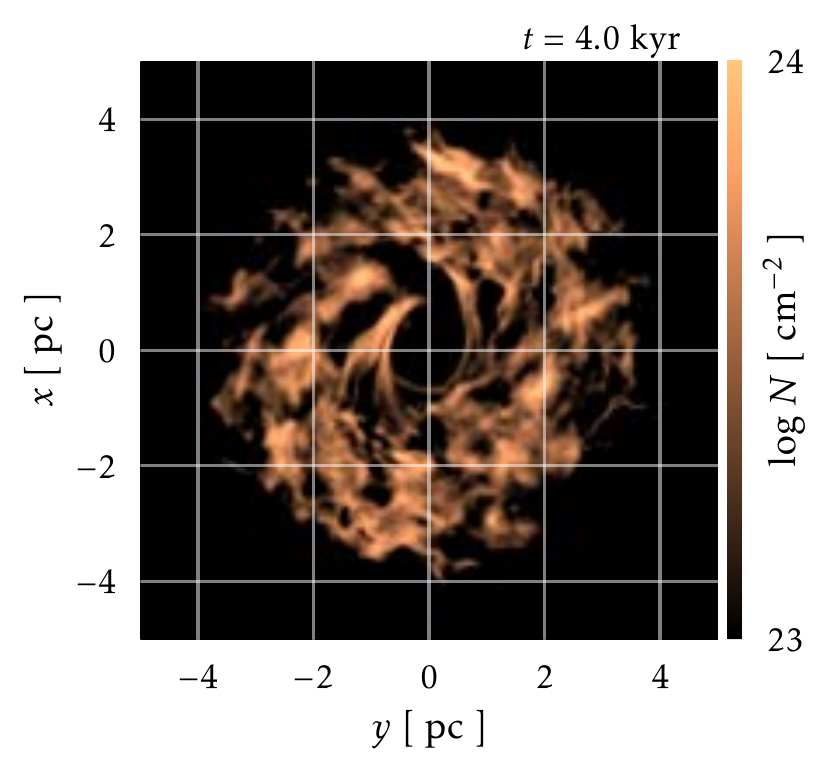} \\[-3.25mm]
    \includegraphics[height=0.27\textwidth]{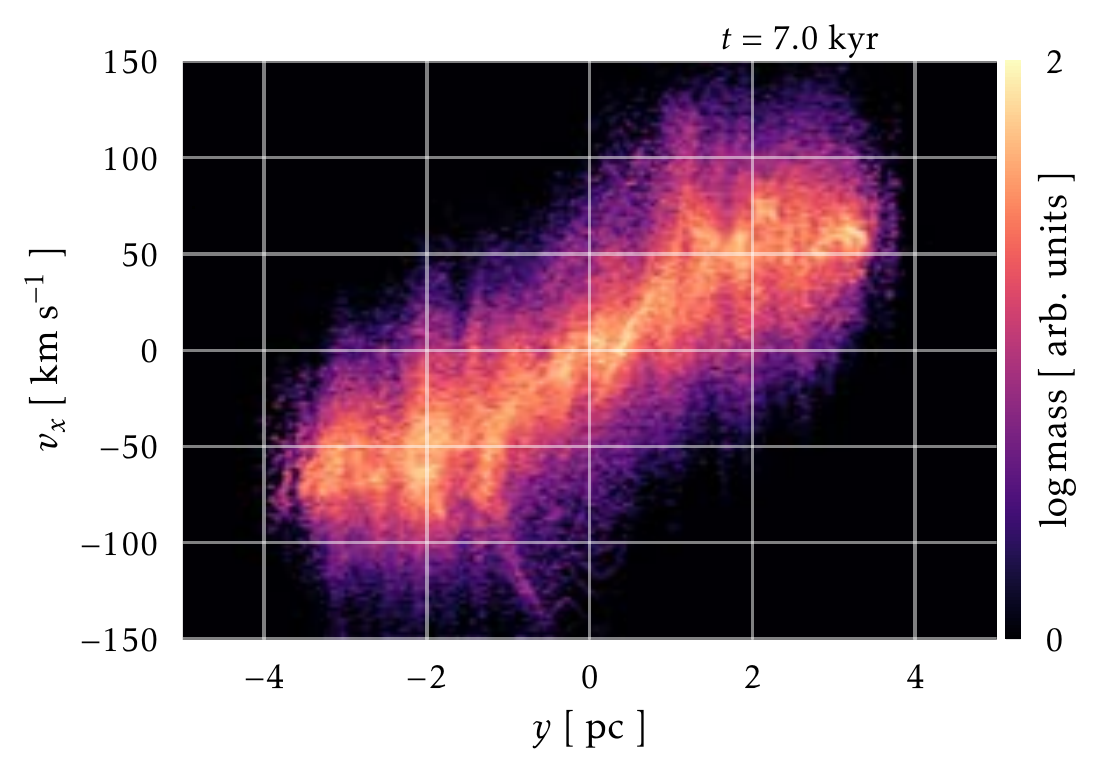}
    \includegraphics[height=0.27\textwidth]{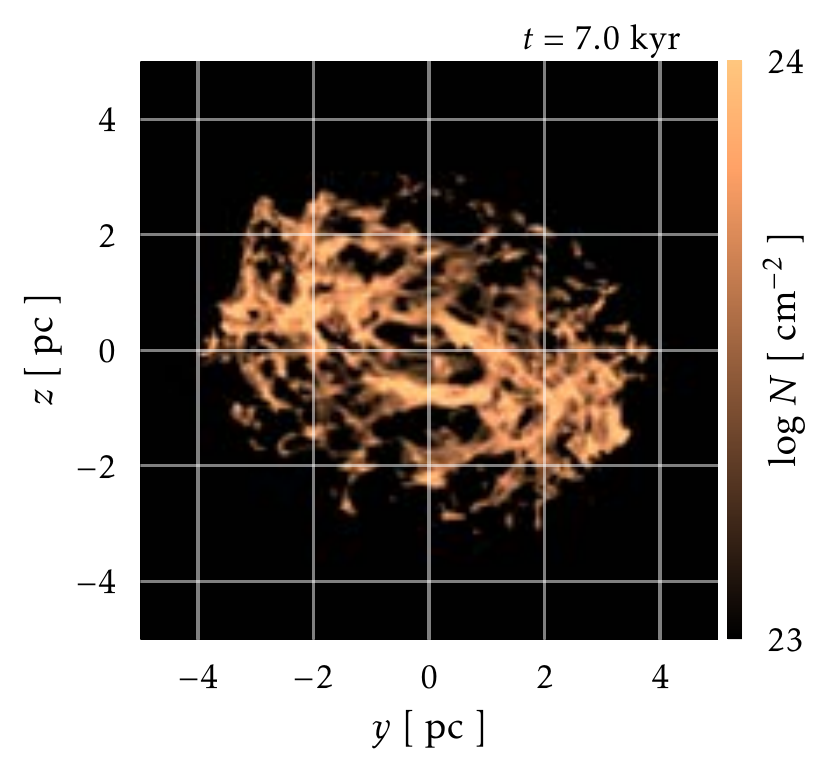}
    \includegraphics[height=0.27\textwidth]{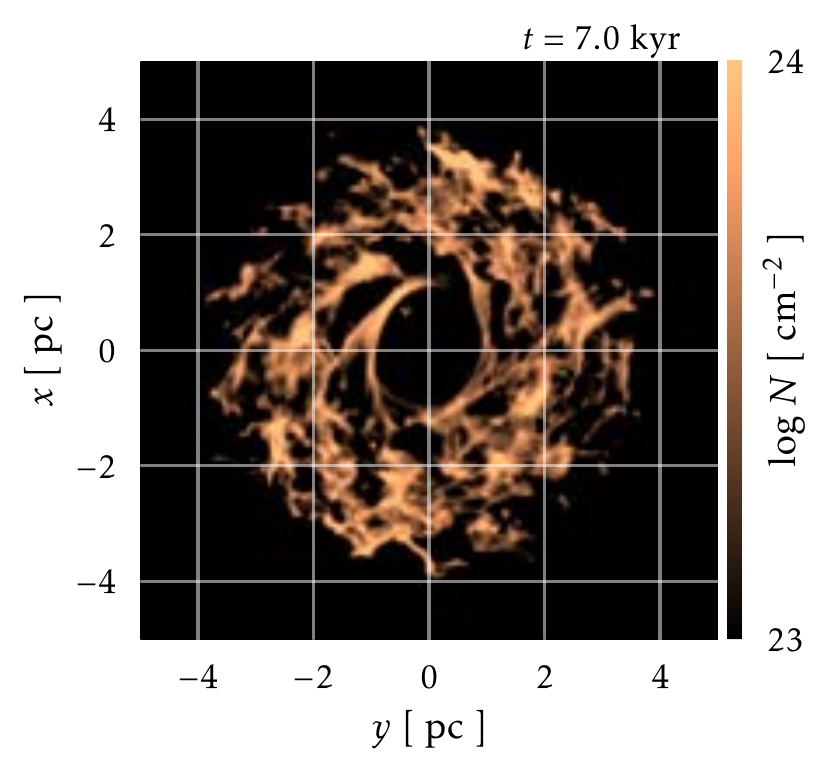} \\[-3.25mm]
    \includegraphics[height=0.27\textwidth]{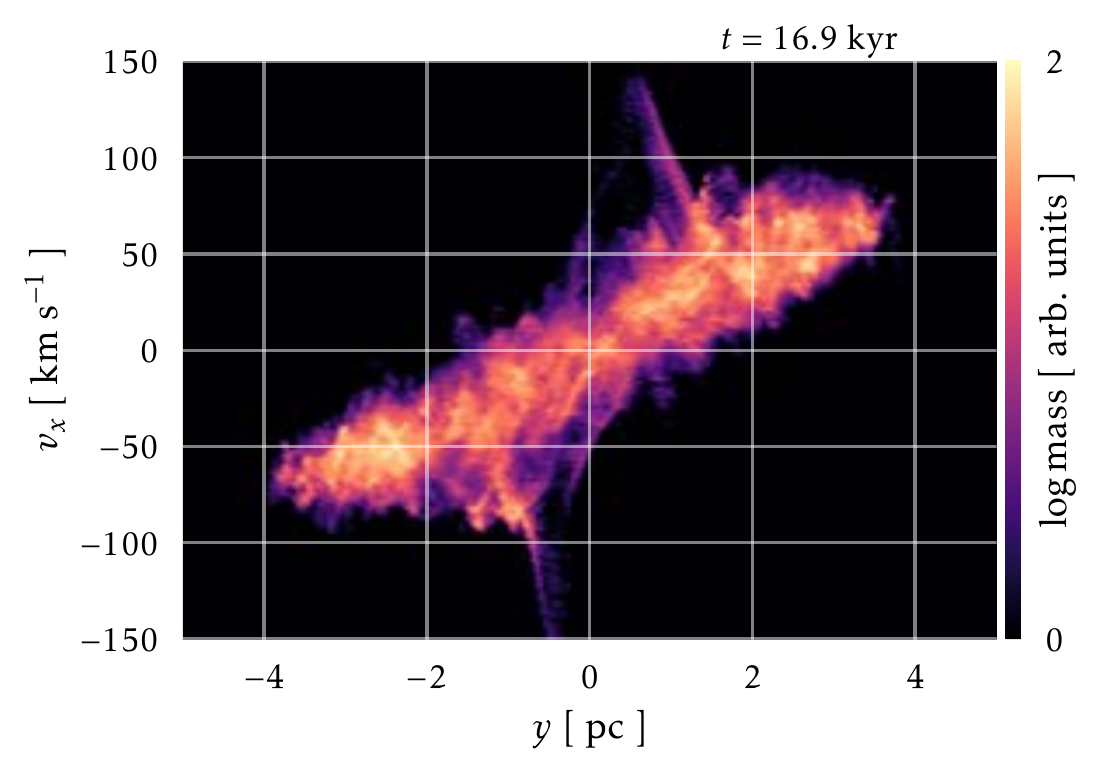}
    \includegraphics[height=0.27\textwidth]{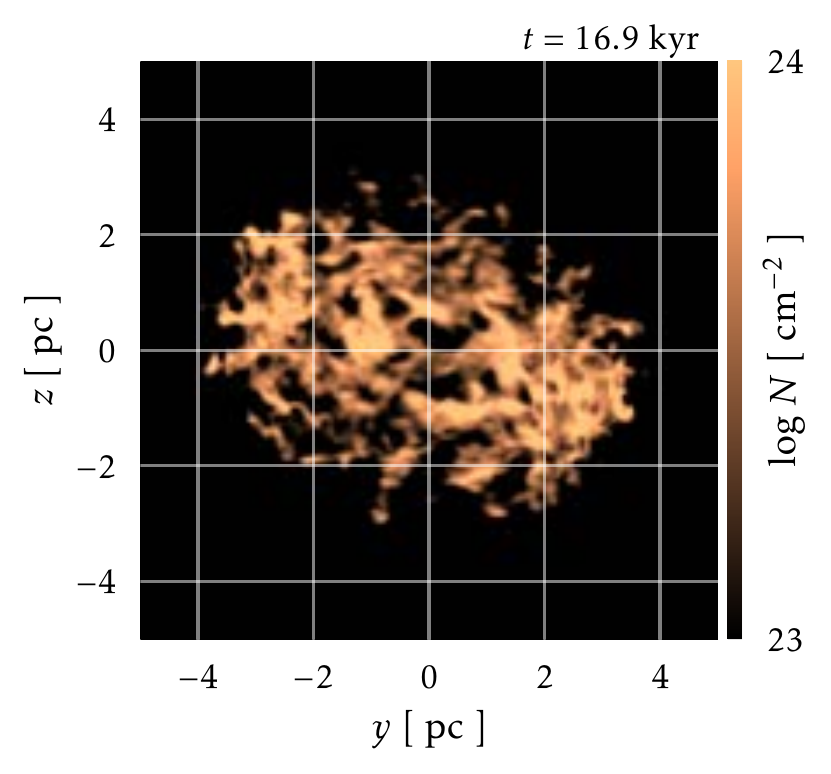}
    \includegraphics[height=0.27\textwidth]{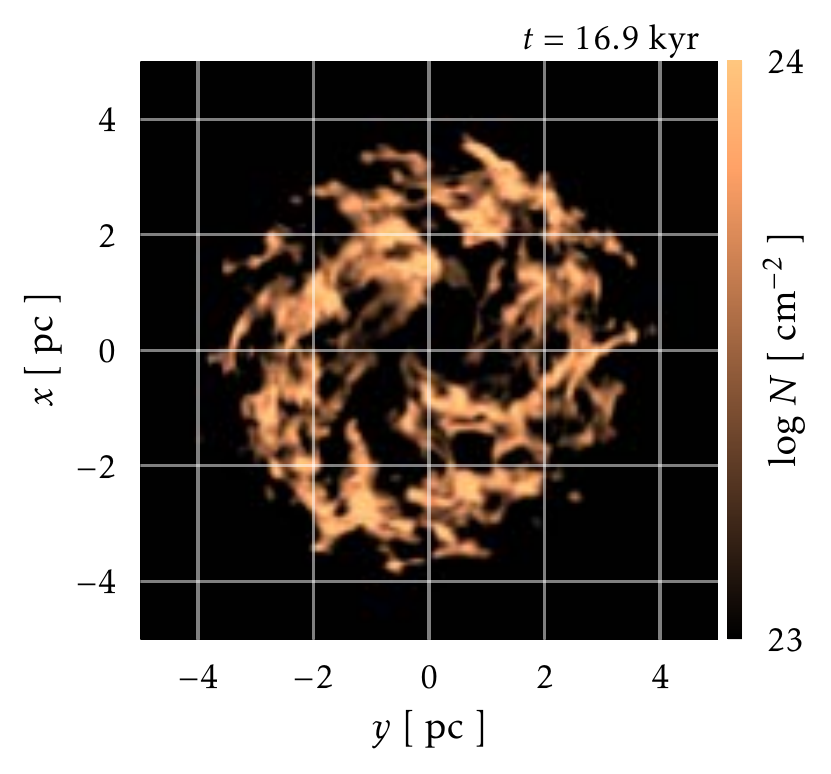}
    \caption{Mass-weighted PVDs along the $y$-axis and the line-of-sight velocity ($v_x$) and column density of the simulations with the clumpy CND. Left panel: PVDs. Middle and right panels: column density of the dense CND gas seen along our line of sight ($x$-axis) and downward ($z$-axis), respectively. Top, middle, and bottom rows are snapshots at $t=4$ kyr, $t=7$ kyr, and $t=17$ kyr (relaxation run), respectively.}\label{f:pc-a6hte-pv}
\end{figure*}

By contrast, the clumpy CND simulations exhibit outcomes of strong jet--CND interactions. The left column of Figure~\ref{f:pc-a6hte-pv} shows results at 4 and 7 kyr, and at the end of the relaxation run. For each we also show the column density of the dense CND gas as seen along $x$ (middle column) and $z$ (right column). Individual clumps are clear in the rotation curve. The CND gas is very fragmented and at $<$2 pc radius is stretched along the rotation curve. Interaction with the jet steepens the gradient in the PVD of each clump, but flattens the full rotation curve \citep[see, e.g.,][]{Mukherjee2018-nd}, a development that is clearest in the snapshot at 7 kyr. Widths and offsets of individual features in velocity space increase with time, but are generally a few tens of \kms. 

At 4 kyr, a sharp, inverted S-shaped antisymmetric feature about (0, 0) appears that persists to $t=7$ kyr. The column density view of the CND shows that this feature is associated with a thin, ring-shaped, strongly sheared and quickly rotating flow near the inner edge of the CND that was compressed by the jet. At 7 kyr, a structure stretches from $v_x=30$\kms{} to $v_x=-140$\kms{}. It is not clear if this is a single spatially continuous structure.

Mixing and recondensation of high density gas clumps as described in Section \ref{s:sim-relax} is also evident in the PVD of the final snapshot from the simulation where the jet-perturbed CND is given time to settle (bottom row of Figure~\ref{f:pc-a6hte-pv}). The gradient across the CND remains shallower than before the jet-ISM interaction, but the clouds are no longer being sheared and compressed by the jet. Therefore, features in the PVD are less stretched as the clouds recondense. Evident at $(y, v_x)\sim\pm(1 \mathrm{pc}, 100$\kms) are two sharp streaks that span $\sim0.5$ pc in $y$ and 70\kms{} in $v_x$; they seem to emanate from clumps. It is unclear if this is a remnant of the inner ring (and associated S-shaped structure in the PVD) in the snapshots while the jet was active because that ring has disappeared.

\section{Discussion}
\subsection{Simulation Caveats and Possible Improvements} \label{s:sim-disc}

These simulations are first steps to explore the effects of the jet on the MW ISM over the past few Myr. We based initial conditions --- including gravitational potential and jet and CND orientations --- on observations and models of the MW. However, the gas distribution just before the most recent jet episode is unknown, as are possible prior jet episodes. The volume filling factor in our initial gas distributions may, therefore, be excessive. As demonstrated with the kiloparsec-scale simulations, however, different filling factors (by adjusting parameter $\rmmm$ in the \citealt{McMillan2017-kl} distribution) can simulate gas distributions resulting from previous jet events. A more extensive survey of filling factors and jet power may then constrain the duty cycle of the MW jet.

Simple simulation physics resulted in encouraging similarities between simulated radio images and the MeerKAT data. We found evidence for several formation scenarios of Anomaly C linked to jet--CND interactions in the GC. There are, however, caveats to consider when comparing models with observations.

First, hydrodynamical interactions between jets and ISM are complex. Outcomes are most sensitive to three parameters that in turn depend on the cloud porosity (filling factor) and scale height of their global distribution: (1) the ratio of jet power to mean ambient density (hot phase+clouds), $\Pjet/\nav$, (2) the mean column density of clouds, $\Nav$, and (3) the duration of jet confinement, $\tconf$. In this paper we have not explored these parameters systematically. However, past work \citep{Wagner2012-cx} shows that jet-blown bubble dynamics are set mainly by $\Pjet/\nav$, and that, for a given $\Pjet/\nav$, the acceleration of clouds embedded within the bubble is proportional to $1/\Nav$. For otherwise identical ISM parameters, $\Nav$ also sets the radio morphology: larger values produce more blobby radio structures due to fewer but more prominently diverging secondary-jet streams, smaller values produce a prominent main jet stream and smoother radio surface brightness variations across the source.

Second, our simulations were designed, and their boundary conditions set, to ensure that the dynamics of the ISM gas in the simulation domain were governed primarily by prolonged interactions with the jet.
However, it is important to check results with adaptive mesh simulations of the full jet to avoid boundary effects and to capture the dynamics of the jet backflow.

Third, the spatial resolution of our grid and numerical diffusion limits the range of density contrasts from gas compression and cooling that we can capture. For the kiloparsec-scale simulations, this limitation does not affect results significantly because global gas dynamics are mostly governed by $\Pjet/\nav$. However, for the CND-scale simulations, numerical diffusion can inhibit formation of a narrow feature such as Anomaly C in the presence of strong shear or compressive forces that smooth out cooling layers to make their surfaces more susceptible to ablation. The CND-scale simulations, therefore, underestimate the mass of narrow filaments, and overestimate the ablation rate and their velocity once entrained. 

Finally, the most important physics omitted here are magnetic fields and thermal conduction, which affect the stability and morphology of clouds embedded in fast flows \citep{Shin2008-mq}. The filaments induced by the fast shear flows of the jet and subsequent radiative cooling may be further supported, and thus prolonged, by magnetic fields. Likewise, magnetic field dependent, anisotropic thermal conduction suppresses the growth of dynamical instabilities at the interfaces between clouds and ambient flow \citep{Orlando2008-pr}.
For the parsec-scale simulations, including magnetic fields will more accurately calculate the radio surface brightness. Including anisotropic conduction may alter the mixing rate of jet plasma and the ISM, which could alter the predicted X-ray surface brightness. For the kiloparsec-scale simulations, magnetic fields and anisotropic conduction will impact strongly the formation of Anomaly~C. Field lines stretched along the jet axis through entrainment and thermal conduction may stabilize dense gas filaments along that axis. Similarly, lateral compression of gas by the jet could enhance the known magnetic field within the CND \citep{2018ApJ...862..150H}, which could in turn facilitate the formation of a linear feature such as Anomaly C by prolonging jet compression.

Early snapshots of our kiloparsec simulations of the second outburst suggest that the jet driving the MeerKAT 1.274 GHz structures may have occurred only 0.1--0.3 Myr ago and flowed into an ambient ISM more porous than that represented by the initial conditions of our simulated second outburst. Perhaps this outburst would form multiple broad streams of radio plasma that emerge from close to the plane of the MW disk and an arc-shaped bow shock within the inner 500 pc. Simulations to explore this scenario will require several-fold higher resolution near there. Close comparison of observational data with these and future refined simulations will constrain the history of the Milky Way's nuclear jet. 

\begin{figure}
    \centering
    \includegraphics[scale=0.44]{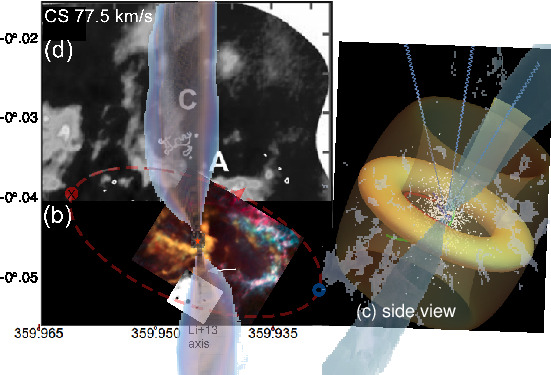}
    \caption{Repeat of Figure~\ref{fig:CND} (b-d), now overlaying the jet particle tracer of the clumpy CND simulation at $t=7$ kyr after the jet flow began. Gas entrained by the jet is not shown. As constructed, the jet projects onto molecular Anomalies C and perhaps A in the north and onto the X-ray jet in the south.
    \label{fig:42ergJet}}
\end{figure}

\subsection{Jet Cocoon Shocks in the North}\label{sec:discussN}

Returning to observations,
Figure~\ref{fig:orbitErrors} shows that the axis perpendicular to the current accretion disk cannot intersect the putative southern jet interaction, but can hit SmR-3 if that nebula is at a $\sim20$ pc radius beyond the far side of the CND.
Conversely, over the allowed jet orientations inferred by asserting its colinearity with the southern jet interaction, the northern counter-jet emerging from the CND within 30\arcdeg\ of the CND's spin axis would impact if SmR-3 is a diffuse extension of the CND 8.6 pc above the Galactic plane at $\sim4.2$ pc radius on either side of the CND.
Because of its sideways tilt to the Galactic plane, the eastern inner edge of the CND is much closer to the jet than is the western (Figure~\ref{fig:CND}(f)) for all orientations.
The simulations emphasize that the jet can alter gas kinematics across the CND and produce localized deviations as large at that of Anomaly A.

The northward jet flow may therefore appear in molecular and ionized gas at the following two plausible jet/ISM interactions.

\subsubsection{At molecular Anomaly C}
The kinematics of Anomaly C described in Section \ref{sec:Njet} are consistent with gas dragged by the jet cocoon from the closest part of the CND, thence receding from us as a cylindrical plume; likewise, the 6\arcdeg\ and $\sim5$\arcsec\ deviations on sky between the western strand and projected jet are naturally explained by projection there (Figure 29).
Starting $\sim40$\arcsec\ beyond the CND, the eastern strand of Anomaly C projected deeper into the CND wall has sufficient optical depth to appear in both CS isotopologues unlike its tenuous western strand.
The velocity trends imply a $\sim10$\kms\ expansion along our sightline of a semi-cylinder inclined away from us to redshift overall by 70-80\kms.
This redshift deprojects to $\la85$\kms\ expanding crosswise or $\ga200$\kms\ flowing axially at constant velocity in the cylinder (Figure~\ref{fig:CND}(e)).
This is not the ``Hubble law" linear acceleration of many protostellar molecular outflows, which attain comparable velocities but have only 1\% of the mass in Anomaly C.

If the jet is now so weak as to have almost stalled, the cocoon is then cool enough to not radiate efficiently so would expand more crosswise to pressurize the ambient ISM and bend flow streak lines to become almost perpendicular to the jet axis \citep[e.g., the pure hydrodynamical simulation of][]{1994ApJ...426..204C}.
That semi-cylinder Anomaly C appears in CH$_3$OH channels points to gas excited by shocks of $\sim10$\kms, which is a plausible velocity for a molecular cloud being shoved aside by an almost adiabatic fossil cocoon.

Anomaly C spans only 20\kms, indicating comparable turbulence throughout a dynamic structure.
The absence of ionized emission and an $\sim10$\kms\ molecular flow suggest a C-shock wherein ambipolar diffusion of neutrals past ions dissipates energy.
However, the poorly constrained local magnetic field $B$, confused CO emission, and its uncertain conversion factor in the GC to H$_2$ density $n_\mathrm{H_2}$ all prevent determination of the local Alfv\'{e}n velocity $V_A \sim1300Bn_\mathrm{H_2}^{-1/2}$ \kms\ that must exceed the flow velocity for a C-shock to form.
Alternatively, the higher 3D velocities of mostly axial motion would imply a more vigorous jet, but whose side entrainment shock would still be slow enough to emit only soft X-rays that are easily attenuated by a molecular screen.
From an optical depth $\tau=3.2$ derived from its CS/C$^{34}$S fluxes, T18 assume LTE to estimate its total $m($H$_2)=10^3~(T_{\rm ex}/200$ K) M$_\odot$ with excitation temperature $T_\mathrm{ex}$ for $n_\mathrm{H_2}\sim10^4$~cm$^{-3}$ from all line constraints. 
The corresponding kinetic energy of Anomaly C $\sim10^{48}(T_\mathrm{ex}/200$ K) erg is at the top end from a single massive protostellar wind, but its molecular mass is at least 100$\times$ larger.

Simulations by \citet{2010ApJ...709...27W} showed that a magnetic field and outflow feedback would prevent a comparable mass from forming many massive stars.
Those authors showed that strong radiative cooling behind a direct jet/ISM impact compresses the flow and ultimately shrinks the bow shock, thereby sweeping up less ambient ISM compared to a weak interaction.
The low observed velocity dispersion of Anomaly C indicates that the jet here is still supersonic, so should end in a noticeable bow shock that expands the cocoon away from the jet to increase entrainment.
Entrainment will eventually make the jet subsonic hence increase turbulence in the surrounding ISM; momentum transfer is most effective at low Mach number through the resulting thin cocoon.
Ambient gas would fill the shroud at the local sound speed in only $\sim4\arcsec/(10$\kms) = 15,000 yr, or perhaps half this time if the Alfv\'{e}n velocity is larger.
\begin{figure}
\centering
\includegraphics[scale=0.34]{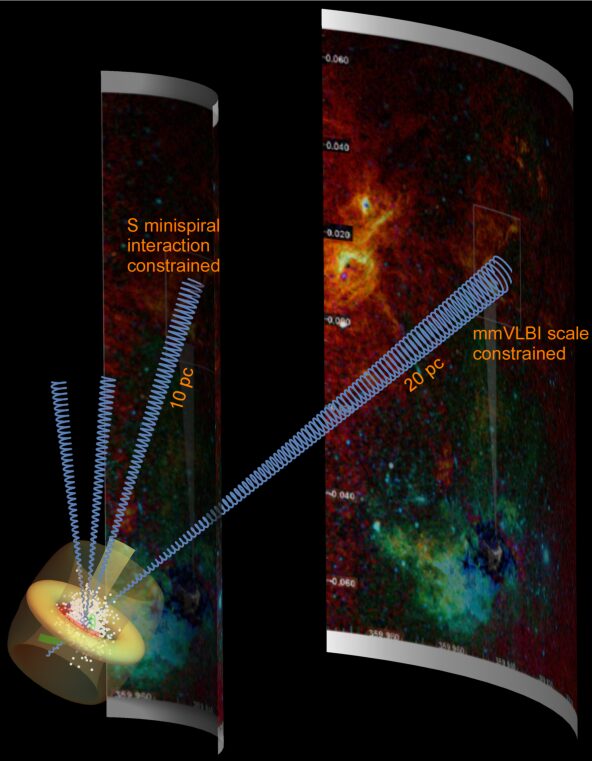}
\caption{Locating spatially nebula SmR-3 (Figure~1) if on the far side of the CND, when constrained by interactions to the S (inner) or at mm-VLBI scale (outer). The alternative nearside placement would be immediately adjacent to the left-hand side of this figure, which we omit for clarity.}
\label{fig:Nposition}
\end{figure}

\subsubsection{At ionized nebula SmR-3}
The Nobeyama space--velocity map of \citet{2016ApJ...831...72H} in Figure~\ref{fig:Nobfig} shows that SmR-3 likely projects on the CS flux peak denoted ``blue ML."
However, the CS line profiles do not split at SmR-3 because their  Figures~6--8 show near-constant 110 \kms\ splitting across at least 35~pc.

In fact, by combining HNCO, N$_2$H$^+$, and HNC data cubes, \citet{2016MNRAS.457.2675H} could track these two distinct kinematical features across at least 1.1\arcdeg\ longitude (160 pc, their Figure~19) to place them $>55$ pc from \sgA\ for all three models that they examined.
This placement was revised by \citet{2020MNRAS.499.4455T} who see in their simulations bi- and trifurcations in the dual bar-fed streams that connect discrete clouds within 30 pc of \sgA\ to the varying distribution of molecular gas at a $\sim100$ pc radius.
Evidently, locating toward the GC a specific cloud like SmR-3 is much more uncertain than just pinning it onto the various patterns of excursions around $x_2$ orbits that were considered by \citet{2016MNRAS.457.2675H}.

Figure \ref{fig:Nposition} shows that the mm-VLBI alignment would place an interacting SmR-3 at $\ga20$ pc from \sgA\ beyond the CND.
Anchored at the southern interaction, a jet directed slightly toward us would place SmR-3 between 10 pc beyond \sgA\ (8.6 pc above the back wall of the CND) and immediately above the front of the CND.
However, the blueshifted Br~$\gamma$ profiles and moderate K-band extinction at its southern edge (Figure~\ref{fig:brgamma}) favor the near side, roughly perpendicular to the spin axis of the current accretion disk inferred from orbiting hot spots \citep{2018A&A...618L..10G}.

P15 discussed SmR-3 as a shock at the edge of the northern X-ray lobe, noting reduced emission downstream at a molecular cloud.
Warm gas in SmR-3 seems to shadow that part of the X-ray lobe.
\pA\ at its southern boundary sums to $2.3\times 10^{37}$ \ergs\ and from \sgA\ spans $\theta(R>9~\textrm{pc})\sim6$\arcdeg = 1.2 pc, a plausible envelope of a Mach $\la10$ jet.
However, what fraction of this emission might be jet powered is unknown until spectral maps constrain the contribution of a wind- or ionization-front from nearby WR \pA\ and X-ray source \object{X174522.6-285844} \citep{2010ApJ...709...27W,2010ApJ...725..188M},
and radiation from CND clusters alone of $10^{39}$ \ergs\
\citep{2007A&A...468..233M}.

\begin{figure*}
\centering
\includegraphics[width=0.9\textwidth]{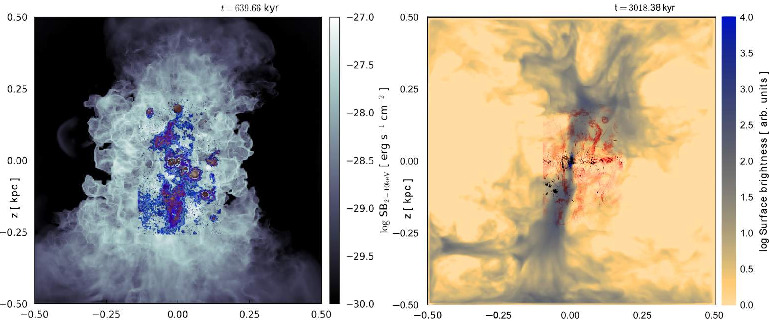}
\caption{Data/simulation comparisons. Hard X-rays (left, from Figure~\ref{f:kpc-xs4} left bottom panel at $t=0.6$ Myr) and radio (right, from Figure~\ref{f:kpc-radio-2} third panel now with inverted color map, at $t=3$ Myr) ray-traces from simulations with the XMM-Newton and MeerKAT images (Figure~2) overlaid. On this scale, the jet flow easily spans the observed structures.}
\label{fig:datacompare}
\end{figure*}

We explored the southern edge of SmR-3 as a planar shock.
The lack of enhanced X-ray emission there bounds its shock velocity $v_{s7}$,
$T_s=3\mu v_s^2/16k=3.2\times10^5 v_\mathrm{s7}^2/x_{ts}$ K $<10^6$ K \citep{1979ApJS...41..555H}.
With $x_{ts}=2.3$ being the postshock concentration of all particles relative to hydrogen in fully ionized gas, we found that $v_{s7}<2.7$.
This is consistent with the axial flow velocity constrained upstream at Anomaly C.
The pre-shock density comes from the radio continuum and \pA\ fluxes that brighten $<0\farcs5$ ($<0.02$ pc) apart (Figure~\ref{fig:Palpha}), hence limit the cooling length (initial compression to recombination) to $d_\mathrm{cool}=2.1\times10^{17}v_\mathrm{s7}^{4.2}/n_0<0.02$ pc here.
Thus a radiant shock must have $n_0>220$ cm$^{-3}$ that would cool gas to 10$^4$ K in $\la75$ yr.
The MAPPINGS code \citep{1993ApJS...88..253S} shows that a planar shock at this velocity and pre-shock density radiates $<1\%$ of its power as \pA\ for any reasonable ambient magnetic field.
Taking spectral slope $\alpha=1$ for $\gamma=10^2-10^4$ of any nonthermal electron distribution of fraction $\eta$, $<0.12$ mJy beam$^{-1}$ from Section \ref{sec:Njet} at boundary 1\farcs5 along our sightline, and equipartition between magnetic energy density and double the electron density to account for cosmic-ray ions, yields $B_0=\eta 3.4\times10^{-4}$ gauss. Compression ratio $<4$ means $\eta\ll1$ and therefore a dynamically unimportant field.

The interval between \sgA\ and SmR-3 on sky is spanned by radial filaments in radio, \pA, and X-ray wave bands (Figures~\ref{fig:ChandraPA}, \ref{fig:radioXray}(e), and \ref{fig:Nobfig}), interspersed with several near-infrared-dark filaments of diminished X-ray emission that converge toward \sgA\ and so are probably at the GC.
The latter resemble dark ``hub-filaments" that characterize star-forming molecular clouds elsewhere in the Galactic disk \citep{2009ApJ...700.1609M} and generally align with the local gravity \citep[e.g.,][]{2020ApJ...905..158W}, here dominated by \sgA\ within the much less massive CND (T18).

\subsection{Streams and Bow Shocks Into the Fermi Bubbles}\label{sec:bs}
Over several hundred parsec scale, shells --- presumably bubbles --- appear.
Many features unrelated to the GC are projected on our sightline, so which of these may not have prosaic origins is receiving attention \citep[e.g., the western half of the GC $\Omega$-Lobe as a foreground H~II region,][]{2020PASJ...72L..10T}.
But certainly Figure~\ref{fig:datacompare} shows that our simulations show radio and X-ray structures at the observed scale. 
Radio flux derived from the $t=3$ Myr simulation reproduces the size and location of bright shells in the MeerKAT image, as well as the $\pm10$ pc scale of the nuclear bilobe shown in Figure~\ref{fig:radioXray}(e) and (b).
Extensive, lower brightness secondary streams extend well beyond the observed structures, while the main stream of this simulation remains well defined as it ascends into the northern Fermi bubble (Figure~\ref{fig:datacompare} right panel).
The $t=0.6$ to 1 Myr X-ray simulation snapshots match the XMM-Newton image; X-rays in snapshots later than $t\sim1$ Myr have faded to invisibility. 
However, if \sgA\ were to enter a period of sustained high luminosity, as fading X-ray fluorescence on nearby nebulae show it once did, then these dynamic X-ray features would brighten anew.

We therefore conclude that a $\sim10^{41}$ \ergs\ jet interacting with the ISM can explain the observed multiwavelength GC structures on both scales.
On 10 kiloparsec scale, the simulations of \citet{2021arXiv210903834M} reproduce the observed boundary of the northern FB when modeled as a forward shock using a jetted flow six times more powerful than ours.

\subsubsection{Comparison with the Seyfert NGC 1068 Outflow}

The GC is plausibly recovering from a Seyfert-level flare from several Myr ago \citep{2019ApJ...886...45B}, augmented by hot-star winds and radiation, with occasional supernovae from the nuclear star cluster.
In prototypical Seyfert \object{NGC 1068}, all of these ionizers are operating on a $<100$ pc radius within its 1 kpc star-forming ring.
Figure \ref{fig:radioXray}(d) highlights remarkable similarities between NGC 1068 radio structures and those in our GC: a core region delineated by elliptical green contours at a $\sim60$ pc radius that coincides with the bundle of vertical nonthermal, strongly polarized radio filaments \citep{2021ApJ...920....6G} in the MW that delineates its radio shells, knots in the compact jet within that have the same scale as the X-ray extension, radio lobes, and GC Lobe in the MW, and blobs at the end of the NGC 1068 jet that distend to become the NE bow-shaped radio lobe at the same scale as the FB in the MW.
The latter shape with adjacent optical and X-ray emission (Figure 32) signifies a radiative, detached bow shock formed when a supersonic jet encounters a rapid decline in ambient pressure.
In the extended narrow-line region of NGC 1068, this deceleration likely occurs near the ``North-East subpeak" bright at 10 $\mu$m \citep{1987ApJ...312..542T} from localized inverse-Compton heating by relativistic electrons impacting warm dusty gas.
HST/STIS spectra show repeated shredding from accelerated clouds up to that point but not beyond \citep{2002ApJ...568..627C} as disintegrating clouds load dust into the outflow.

The jet at $<100$ pc in pink contours in Figure~2(d) is not colinear with the bow-shaped region beyond; it twists on the sky \citep[Figure~2(d),][]{2004ApJ...613..794G}.
Indeed Figure~18 of \citet{2018MNRAS.479.5544M} from simulations similar to ours shows large deviations from the launch axis of hot bubbles at the ends of a decelerating jet.
In the GC, the CND provides the first impact site of a jet even if launched fairly close to the Galactic pole.

\begin{figure}
\centering
\includegraphics[scale=0.3]{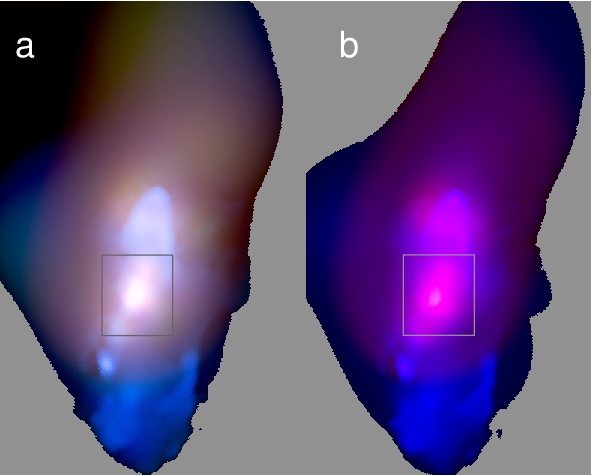}
\caption{The inner $7\times15$ arcsec$^2$ of NGC 1068 (rotated 45\arcdeg\ west through north) shows similar structures to our GC; the box delineates the region contoured in Figure~\ref{fig:radioXray}(d). Part (a) shows in blue the radio jet from 4.9 GHz continuum VLA \citep[Figure 1]{1987ApJ...319..105W}, in red Chandra 0.25--3 keV X-rays \citep[Figure 3]{2001ApJ...556....6Y}, and in (part a only) green optical-band emission from HST blurred to the resolution of the X-rays. 
Note the ``blobby" nature of the radio jet before it flares into bow-shaped lobes, and how to the north X-rays brighten along its bottom boundary, whereas to south they are extinguished by the dense ISM.
Jet orientation has been constrained by ionized gas kinematics from HST and ground-based spectra \citep[e.g.,][]{2002ApJ...568..627C}.
\label{fig:NGC1068xray}}
\end{figure}

\subsection{Jet Visibility}\label{sec:visible}

A jet appears by particle cooling, compression/re-acceleration at shock fronts, recollimation, and reconnection of magnetic flux tubes.
Such opportunities seem scarce in the GC filled today with a hot flow that has banished clouds to a sheath at least 2.5 pc and perhaps 17 pc distant.
Our simulations in Section \ref{sec:sims} show that the kinematics of a structure as extensive as the broad Polar Arc would require prolonged exposure to a weak jet because a powerful one pounds through the inhomogenous ISM, enduring many cloud impacts but emerging still fairly collimated within 0.1 Myr.
By contrast, a low-power jet almost stalls to blow a bubble in the disk, which eventually becomes a hot vertical channel as the jet cocoon backflows \citep[e.g.,][]{2018MNRAS.479.5544M}.
Most gas barely accelerates radially.
Interactions strengthen as the launch axis is declined toward the disk plane, but the $\la30$\arcdeg\ angular displacement of the current jet from the Galactic polar axis argues for prolonged action of a low-power jet.

All interactions proposed here are localized, not broad and windy, impacts on dense gas.
Therefore, the observed patterns are consistent with a momentum-driven brightening arising from the ram pressure of the jet cocoon around the low-power outflow.
However, the radial pattern of dark filaments, X-ray, and radio emission in Figure~\ref{fig:ChandraPA} suggest a broader wind outflow around the jet.

LMB13 reproduced the spectral slopes of the southern jet from X-ray strength to radio upper detection upper limit with a single-zone synchrotron model that \citet{2019ApJ...875...44Z} then applied along the entire feature.
The latter authors showed the need for a strong magnetic field to seed inverse-Compton emission from hot dust, whereupon the resulting radio synchrotron would be very strong.
However, the X-ray emitting jet is not a radio source, so they thereby obtained a brief X-ray synchrotron cooling time $1.2(B/ \mathrm{1~mG})^{-1.5}(E/\mathrm{5~keV})^{-0.5}$ yr. 

Thus, the jet appears today solely by its interaction with a minispiral arm; the counter-jet at similar radius would project behind most allowed orbits of the Western Arc, so cool gas may be sparse.
Adding post-shock cooling at the southern interaction totals $\sim10^{33}$ \ergs\ from a jet having much more power. 
LMB13 constrained the power of the jet today as its particles ram through the southern shock $GM_\mathrm{BH}\mu(T)m_H n_f/R\times\pi(R\theta/2)^2v_f\approx 2\times 10^{37} (R/0.5\textrm{pc})(\theta/25\arcdeg)^2(n_f/10^4~\textrm{cm}^{-3}) (v_f/200$ \kms) \ergs\ with $R$ the distance between the shock and \sgA\ of mass $M_\mathrm{BH}$, $\mu(T)=1.2$ the mean molecular weight in warm gas, $n_f$ and $v_f$ the density and Keplerian orbital velocity of the intersected gas in the Eastern Arm, and jet opening angle $\theta\sim25$\arcdeg\ set by LMB13 to span the uncertain spread of the cocoon at that location.
Our simulations therefore show that the current jet can have faded by four orders of magnitude since it formed the observed large-scale structures several Myr ago.

\section{Summary and Conclusions}
Allowed jet orientations at mm-VLBI and VLTI scales are converging.
Beyond, this paper has used ALMA spectral, and Chandra, HST, and JVLA imaging archival datasets with new SOAR telescope near-infrared spectra to propose several signatures of a weak jet near PA\ 121\arcdeg:

\begin{enumerate}
\item LMB13's X-ray ``streak'' with a nonthermal spectrum south of \sgA\ and the associated ``Seagull Nebula" shock feature just upstream in one arm of the nuclear minispiral. 
In this hemisphere, no other jet-like features appear in molecular or ionized gas.
\item Almost diametrically opposite on sky is Anomaly C, an elongated kinematical ``tickle" in molecular gas whose properties T18 derived from its relative surface brightening in several diagnostic emission lines:
    weakly shocked molecular gas at $\la10^4$ cm$^{-3}$ with $m(\mathrm{H}_2)=10^3 (T_\mathrm{ex}/200$ K) \Msun, kinetic energy $>6\times10^{49}$~erg, and velocity dispersion $\sim10$\kms. 
    Uncertain minispiral orbits permit a counter-jet inclined \ranA\ to our sightline, thus would deviate substantially from the axis of the current accretion disk.
    Our perspective on the jet projects its interaction with the ISM nearer to the far wall of the dense, tilted CND, with much of the allowed inclination range indicating mostly axial outflow at $\la200$\kms\ and with almost no turbulent broadening; if its motion is mostly transverse expansion then dense gas nearby is being compressed by $>10$\kms.
\item A possible jet/molecular cloud interaction 3\farcm6 to the north appears in \pA\ and 5.5 GHz continuum at the southern boundary of SmR-3 of Z16.
    Here, a wedge of constant brightness suggests a jet cocoon that has not expanded from a 3\arcdeg-6\arcdeg\ opening angle after 10 pc of travel.
    It has comparable luminosity to the southern jet/ISM interaction yet spans only 6\% of that structure's solid angle.
    Our near-infrared spectra show that the structure is blueshifted by 70\kms, which places it 40 \kms\ from the blue component of the ML.
It seems to shadow the northern X-ray lobe and has moderate K-band extinction for the GC.
    Locating this putative ISM interaction on the nearside of the CND would orient the jet far from the spin axis of the observed accretion disk.

\end{enumerate}

We performed relativistic, hydrodynamic simulations of a jet of power $\Pjet=10^{41}$ \ergs{}. The jet expels most gas in the central kiloparsec within 5--10 Myr as it propagates through the height of the MW disk, consistent with the observed H~I hole there. Imposing a second outburst through a partially refilled central region cleared that gas within 3 Myr, although jet interactions with small clouds persisted to 10 Myr to broaden the jet into several streams. As a result, bright regions of the synthetic radio surface brightness images from later times of this simulation match the lateral extent to the MeerKAT 1.28 GHz image. Synthetic X-ray images made from our simulations exhibit cavities and blobby structures at later times that anti-correlate spatially with the radio surface brightness as evident in Figure~1. Interactions of jet streams with ISM clouds drive radiative shocks into the outer cloud layers that appear in soft X-rays.
Strong head-on interactions can sometimes form hot spots of hard X-rays.  Clouds that are ablated and carried outward with the jet streams often form filamentary or cometary head-tail structures. Some jet-cloud interactions deform clouds to resemble SmR-3 seen in \pA\ and JVLA 5.5 GHz. 

We likewise performed simulations on 10 pc scales, where the jet of identical power interacts with the torus-like CND. We studied a smooth CND and two cases of a clumpy one. For each simulation, the jet was switched off after prolonged interaction with the CND to follow the subsequent relaxation. Each simulation revealed a possible formation mechanism for Anomaly C:
\begin{enumerate}
\setcounter{enumi}{3}
    \item {\it Smooth CND}: The CND itself cannot be penetrated by jet plasma, but gas ablated from its impacted surface forms narrow filaments that align and persist with the active jet. The densities may be too low and velocities too high for these features to be associated with Anomaly C.
    \item {\it Clumpy CND -- fragmentation}: Jet plasma perturbs gas inside the CND to trigger its compression, runaway radiative cooling, and collapse. The collapsed filaments tend to align parallel or perpendicular to the jet and become more prominent later in the simulations. While their densities and velocities resemble the properties of Anomaly C, their lifetimes are uncertain but likely persist due to their high density. Filaments collapsing in the plane of the CND and pushed to one side by the jet may counter-rotate like Anomaly A.
    \item {\it Clumpy CND -- cloud lift-up}: When clouds are in the jet path or are subject to strong pressure gradients in the jet-blown bubble, they accelerate in bulk to form an elongated cometary head-tail structure of density and velocity intermediate to the two features described above. They appear early in the interaction but persist only while the clump is accelerated (a few kyr).
    \item {\it Relaxation}: After the jet has switched off, the CND gas and any filaments formed through the jet--CND interactions tend to expand as the surrounding thermal and turbulent pressure drops suddenly. As the turbulence decays, radiative cooling of the mixed gas reforms some filaments.
\end{enumerate}

Some of the structures listed appear in the PVD of our simulations as sharp streak deviations of tens of \kms\ from the mean rotation curve. In simulations with the clumpy CND, clumps within 2 pc radius tend to stretch along the rotation curve. Interactions with the jet also cause the gradients of clumps in the PVD to steepen and those of the entire CND to flatten slightly.

While detailed comparison with the observed PVD is left for future work, we are encouraged that a $10^{41}$~\ergs\ jet interacting with an inhomogeneous ISM explains both kiloparsec and parsec-scale GC structures. Note that the initial MW mass distribution (as modeled by the \citealt{McMillan2017-kl} profiles) and the inclination of jet to CND and to the MW disk are very much specified here to examine the case for a jet during the past 1--10 Myr. Due to the tilt of the CND, our results are robust to jet launch angles. These simulations are merely first steps in modeling past interactions of the MW ISM with a jet and their relics evident today. More detailed quantitative comparisons using MHD simulations at higher resolutions will be presented in a later paper.

The regions noted around the CND and our simulations on both scales plausibly delineate an ongoing, albeit currently weak, more collimated outflow within a broader double-lobed molecular and X-ray outflow anchored on the CND.
\sgA\ is dim today, but was ten thousand times brighter only a few centuries ago.
Sustained brightening by only a hundredfold from today could replenish the jet sufficiently to glow along its entire path a few decades thereafter.

\software{Astropy \citep{astropy:2013, astropy:2018} v4; CIAO \citep{2006SPIE.6270E..1VF} v4.11; PLUTO \citep{2012ApJS..198....7M} v4.3; Mathematica v12}

\acknowledgements
This paper uses ALMA archival data ADS/JAO.ALMA\#2012.1.00080.S. ALMA is a partnership of ESO (representing its member states), NSF (USA) and NINS (Japan), together with NRC (Canada), MOST and ASIAA (Taiwan), and KASI (Republic of Korea), in cooperation with the Republic of Chile. The Joint ALMA Observatory is operated by ESO, AUI/NRAO, and NAOJ.
The National Radio Astronomy Observatory is a facility of the National Science Foundation operated under cooperative agreement by Associated Universities, Inc.
JSPS KAKENHI Grant Number 19K03862 supported our numerical work.
The 5.5 GHz image obtained with the Karl G. Jansky Very Large Array (JVLA) was downloaded with thanks to Dr.~Zhao from \url{lweb.cfa.harvard.edu/~jzhao/GC/sgra/Feedbk2015apj/SGRAU.FITS}
The scientific results reported in this article are based in part on observations made by the Chandra X-ray Observatory and published previously in the cited articles.
HST data presented in this paper were obtained from the Mikulski Archive for Space Telescopes (MAST) at the Space Telescope Science Institute. The specific observations analyzed can be accessed via \dataset[doi:10.17909/T9JP40]. STScI is operated by the Association of Universities for Research in Astronomy, Inc., under NASA contract NAS5-26555. Support to MAST for these data is provided by the NASA Office of Space Science via grant NAG5-7584 and by other grants and contracts.
Numerical calculations were performed on the National Computational Infrastructure (NCI) facility at ANU. 
Infrared spectra were obtained at the Southern Astrophysical Research (SOAR) telescope, which is a joint project of the Minist\'{e}rio da Ci\^{e}ncia, Tecnologia e Inova\c{c}\~{o}es (MCTI/LNA) do Brasil, the US National Science Foundation's NOIRLab, the University of North Carolina at Chapel Hill (UNC), and Michigan State University (MSU). We thank Dr.~Sean Points (NOIRLab) for advice on tSpec and the referee for a careful review.
\facilities{ALMA, JVLA, HST(NICMOS), Chandra, SOAR(tSpec), Nobeyama 45m, NCI} 

\appendix
\section{Simulations} \label{s:simulations}
\subsection{Simulation Code and Setup Common to All Runs}

All simulations here were performed with the PLUTO code \citep{2012ApJS..198....7M} using its relativistic-hydrodynamic module. The warm or cold phase, the hot phase, and the relativistic jet are treated as a single fluid. All simulations included a static gravitational potential and employed a tabulated radiative cooling generated by MAPPINGS V \citep{Sutherland2017-ep}. Radiative cooling is crucial here because the clouds cool very quickly compared to the simulation time and, being shocked, evolve very differently to adiabatic clouds. They fragment through thermal instabilities but form dense cores and long filaments \citep{Sutherland2007-rn, Cooper2009-ou, Wagner2011-rh, Wagner2012-cx, Mukherjee2016-wf, 2018MNRAS.479.5544M}. 

The uniform Cartesian grid in the cubical domain in all simulations contained $256^3$ cells. The right-handed coordinate system is always oriented such that the negative $x$-axis points toward the Earth observer, that is, the positive $x$-axis points from the observer toward \sgA, and the positive $z$-axis points northward. The simulation domain was always centered at the coordinate origin where a jet inlet was placed using a region of fixed domain-interior cells. This inlet is described in detail in \citet{2018MNRAS.479.5544M}. The jet is bidirectional, jet and counter-jet are rotationally symmetric, and the orientation and opening angle of the jet inlet can be set arbitrarily. The jet orientation is a crucial parameter for this study, whereas a modest variation in the opening angle has little effect on the outcome because the over-pressured jet quickly expands then recollimates regardless of the precise opening angle used. Within the jet inlet, the (primitive) cell values are set to constant precomputed jet pressure, density, and velocity values. 

The primary jet parameter governing the evolution of the jet as it interacts with the ISM is its power, $\Pjet$. This together with the jet bulk Lorentz factor $\Gamma$, and the ratio of rest mass energy density to pressure of the jet, $\chi$, determine how heavy and how over-pressured the jet is. The relationship between these quantities and the jet pressure and density, given a cross-sectional area of the jet, can be found in \citet{Sutherland2007-rn} and \citet{Wagner2011-rh}. The jet base is always resolved by at least 12 cells to capture internal shocks properly.

All simulations ran over timescales $\sim 50\times$ longer than the jet crossing time through the domain. Our setup neglects the effect of jet backflows after the jet has reached the end of the domain, but we aimed to maximize spatial resolution for following the evolution of the gas. For the regions considered in this work, the effect of secondary-jet streams percolating through the ISM and the heating and pressurization laterally to the main stream are likely more important than the effects of the backflow, and are likewise captured at maximum spatial resolution in this setup.

Custom outflow boundary conditions that modify PLUTO standard ``outflow'' were imposed on all six surfaces of the cubical domain walls: in addition to first setting the values of the ghost cells to those of the first cell interior to the boundary surface, the velocity component normal to the boundary surface was set to always face outward. This reduced the amount of spurious inflow from the boundaries caused by turbulent eddies that skim the boundary surfaces, and enabled us to run our simulations for much longer than with the standard ``outflow'' boundary conditions. 

We used a third-order Runge-Kutta time integrator and the piecewise parabolic method \citep{Colella1984-ba} of PLUTO in dimensionally unsplit mode, and the relativistic HLLC Riemann solver \citep{2005MNRAS.364..126M}.

In the following we detail the numerical setup for the two scales on which the simulations were conducted.

\subsection{Setup for the Kiloparsec-scale Simulations} \label{s:kpc-sim}

To capture the full extent of the MeerKAT 1.274 GHz data and the 1.5--2.6 keV XMM-Newton X-ray data with these simulations, we chose a domain size of 1 kpc on a side (4 pc cells).

We employed a two-phase turbulent ISM for our simulations, similar to those used by \citet{2018MNRAS.479.5544M}. We used the model of the MW radial and vertical gas distributions by \citet{McMillan2017-kl} to set up the fixed gravitational potential and radial and vertical mean gas density profiles of the two-phase ISM. The isothermal hot-phase gas at $10^{7}$ K and central density of $0.003\ppcc$ is initially in hydrostatic equilibrium with the gravitational potential. Embedded within, and following the density profile of the H~I disk in the McMillan distribution, are clouds that rotate with Keplerian velocity in the gravitational potential and possess a Gaussian random, but spatially correlated, random velocity dispersion of 10\kms{}. The length scale of $\sim120$ pc over which the velocities are correlated is the same as that for the density distribution of the fractal clouds. The clouds are initially in pressure equilibrium with the hot phase, and the initial porosity of the clouds arises because cloud cells are only initialized if colder than $3\times10^{4}$ K. More details can be found in the numerical setup sections in \citet{Sutherland2007-rn} and \citet{2018MNRAS.479.5544M}.

McMillan profile parameter $\rmmm$ controls the length scale of the central hole in the the profiles of the H~I and molecular disks. In the fiducial models of McMillan, the central kiloparsec is largely devoid of H~I and molecular gas ($\rmmm = 4$ kpc), compared to the disk beyond a kiloparsec as observed. When the MW jet became active, the central kiloparsec was likely filled with denser gas, which was then blown away by the jet. Therefore we adopted the fiducial McMillan model and changed only the length scale of the central H~I hole. In the first simulation we set $\rmmm = 0$ (no hole), and in another, representing a second outburst, we set $\rmmm = 0.3$ kpc (partly filled central hole).
The bidirectional jet was oriented such that the N jet was tilted 12.5\arcdeg\ west and 45\arcdeg\ away from us. Unknown are the jet power, other jet parameters, and the typical size scale of clouds in the Galactic disk when the jet became active. However, they can be constrained by examining the efficiency with which the jet disperses clouds and by comparing the simulated radio and X-ray morphologies with present-day observations. The fiducial jet power used throughout this study is $\Pjet=10^{41}$ \ergs{}, although we found similar results for jets an order of magnitude more or less powerful.
Over 3 Myr this jet alone would inject 10\% of the canonical 10$^{56}$ erg discussed in prior studies of the Fermi bubbles, but successfully reproduces structures on smaller scales as discussed in this paper.

The simulations were performed over the freefall time of the interstellar gas in the simulation domain. 

\subsection{Setup for the GC-scale Simulations}

Our simulations on GC scales explored the interactions of the jet with the CND. As for the kiloparsec-scale simulations described above, the initial conditions used for the CND do not represent its gas distribution observed today. We do not know what the gas distribution and properties were when the jet became active, and on the GC scales we do not have a well-constrained model, such as the McMillan profiles for the MW on kiloparsec scales. We therefore examined the jet--gas interactions with several different setups that approximate a gas-rich, toroidal CND. 

The domain size is always 10 pc a side (using 0.01 pc = 0\farcs25 cells) to capture the extent of the ALMA data described in Section~\ref{sec:allN}, including Anomaly C. We included only the supermassive BH + nuclear star cluster as a static gravitational potential of mass $4.1\times10^{6}$\Msun{}, because the domain is contained within its sphere of influence. The CND is oriented such that the northern part of its symmetry axis is tilted by 30\arcdeg\ toward the back and 20\arcdeg\ toward the west. 

The bidirectional jet is oriented such that the northern jet is tilted toward the west by 3\arcdeg, and away from us by 35\arcdeg. It now has $\Gamma = 5$ and $\chi = 1.6$.

The simulations were conducted over timescales comparable to the orbital period at 1 pc and approximately a quarter (an eighth) of the orbital period at 2 pc (3 pc).

Unless a sink is present, the steep gravitational potential near \sgA\ can quickly accumulate and heat gas with effects on the CND dynamics. We therefore created a region of cells in the flanks of the jet to gently extract accreting gas from the domain so that the dynamical evolution of the CND gas results solely from the interaction with the jet.

The two different initial conditions are:

\begin{enumerate}
    \item[a)] {\it Smooth CND.} The CND in the GC was approximated by a cylindrically symmetric, strongly flared disk structure clipped at a height of 2 pc and rotating with a Keplerian velocity in the BH gravitational potential. The density profile follows Equation 2 in \citet{2018MNRAS.479.5544M} with the dimensionless parameter that controls the degree of flaring $\epsilon=0.98$. No random velocity component was imposed on the CND in this setup. 
    \item[b)] {\it Clumpy CND.} The smooth CND gas density profile in Setup (a) is apodized by pre-generated fractal clouds, much like in the setup for the kiloparsec-scale simulations described in Section~\ref{s:kpc-sim}. The clouds also possessed a velocity dispersion of 10\kms{}. The clumpy CND was then allowed to relax over approximately a quarter of a rotation before the jet was injected. Relaxation settles the gas into a state in which the turbulent velocity field become consistent with the turbulent density field, thereby establishing a Kolmogorov spectrum \citep{2018MNRAS.479.5544M}.
\end{enumerate}
\bibliography{MWbib}{}
\bibliographystyle{aasjournal}
\end{document}